\documentclass[fleqn,usenatbib,useAMS]{mnras}

\pdfoutput=1 
\usepackage{amssymb}
\usepackage{amstext}
\usepackage{amsmath}
\usepackage{epsf}
\usepackage{graphicx}
\usepackage{graphics}
\usepackage{epsfig}
\usepackage{wrapfig}
\usepackage{listings}
\usepackage{float}
\usepackage{esint}
\usepackage{natbib}
\usepackage{ifthen}
\usepackage{hyperref}
\usepackage{verbatim}

\usepackage{xcolor}
\usepackage[normalem]{ulem} 
\newcommand\hl{\bgroup\markoverwith
  {\textcolor{yellow}{\rule[-.5ex]{2pt}{2.5ex}}}\ULon}

\newcommand\rhl{\bgroup\markoverwith
  {\textcolor{orange}{\rule[-.5ex]{2pt}{2.5ex}}}\ULon}
  
 \newcommand\ghl{\bgroup\markoverwith
  {\textcolor{green}{\rule[-.5ex]{2pt}{2.5ex}}}\ULon} 
\begin{document}
\label{firstpage}
\pagerange{\pageref{firstpage}--\pageref{lastpage}}

\title[Dark matter haloes: a multistream view]{Dark matter haloes: a multistream view}

\author[Ramachandra \& Shandarin]
	{Nesar S. Ramachandra, \thanks{E-mail: nesar@ku.edu} 
	Sergei F. Shandarin, \\
	Department of Physics and Astronomy, University of Kansas, Lawrence, KS 66045}

\maketitle
\begin{abstract}

Mysterious dark matter constitutes about 85\% of all mass in the Universe. Clustering of dark matter plays the dominant role in the formation of all observed structures on scales from a fraction to a few hundreds of Mega-parsecs. Galaxies play a role of lights illuminating these structures so they can be observed. The observations in the last several decades have unveiled opulent geometry of these structures currently known as the cosmic web. Haloes are the highest concentrations of dark matter and host luminous galaxies. Currently the most accurate modeling of dark matter haloes is achieved in cosmological N-body simulations. Identifying the haloes from the distribution of particles in N-body simulations is one of the problems attracting both considerable interest and efforts. We propose a novel framework for detecting potential dark matter haloes using the field unique for dark matter  -- multistream field. The multistream field emerges at the nonlinear stage of the growth of perturbations because the dark matter is collisionless. Counting the number of velocity streams in gravitational collapses supplements our knowledge of spatial clustering. We assume that the virialized haloes have convex boundaries. Closed and convex regions of the multistream field are hence isolated by imposing a positivity condition on all three eigenvalues of the Hessian estimated on the smoothed multistream field. In a single-scale analysis of high multistream field resolution and low softening length, the halo substructures with local multistream maxima are isolated as individual halo sites.


\end{abstract}

\begin{keywords}
methods: numerical -- cosmology: theory -- dark matter -- large-scale structure of Universe 
\end{keywords}

\begingroup
\let\clearpage\relax
\endgroup
\newpage

\section{Introduction} 
\label{sec:intro}

The web-like distribution of matter initially revealed by redshift surveys (with less than 300 galaxies by \citealt{Gregory1978} and around 1000 galaxies by \citealt{DeLapparent1986}) and numerical modeling (using N-body simulations of around 30000 particles by \citealt{Shandarin1983} and \citealt{Klypin1983a}) pioneered morphological investigations of the cosmic web structures (see \citealt{Bond1996}, also reviews by \citealt{Shandarin1989} and \citealt{Weygaert2008}). Detailed mapping of the Universe has crossed three million objects today, by catalogues such as the Sloan Digital Sky Survey (SDSS; \citealt{Albareti2016}). The upcoming Large Synoptic Survey Telescope (LSST; \citealt{lsst2009}) is expected to probe the nature of dark matter using several billion galaxies. On the other hand, cosmological simulations have improved immensely in several aspects -- numerical techniques, parallelization schemes, inclusion of various physical processes, volume and resolution (some of these developments are summarized in \cite{Bertschinger1998} and \cite{Bagla1997}. Modern state-of-the-art simulations like the Illustris Project \cite{Vogelsberger2014}, the EAGLE project \cite{Schaye2015} and Q-Continuum \cite{Heitmann2015} use more than a billion dark matter particles. Finally, the ever improving data analysis techniques have resulted in new and sophisticated density estimators, geometrical and topological indicators. A plethora of algorithms for identifying and characterizing dark matter structures have emerged in last two decades (a summary on cosmological data analysis is highlighted in \citealt{Weygaert2009a}). Considering all these improvements, it is worth noting that the proto-structures detected in the modern simulations are qualitatively similar to the quasi-linear description of clustering by Zeldovich Approximation (ZA; \citealt{Zeldovich1970}). Location and properties of these structures, i.e., the voids, walls, filaments and haloes maybe inconsistent across different structure finding algorithms, but that is primarily due to varied definitions.

Most of structure finders are halo finders only and majority of them are stemmed from three underlying algorithms. One of them is the SO (Spherical Over-density) halo finder that defines halos  as spherical regions whose mass density exceeds the mean density by a specified factor \citep{Press1974}. 
Another is the FOF  (Friends-of-Friends) halo finder describing haloes  as the groups of particles separated less than a specified linking length often chosen as 0.2 times the mean particle separation \citep{Davis1985}. 
The FOF can be also used for identifying filaments and walls/pancakes by increasing the linking length (\citealt{Zeldovich1982}, \citealt{Shandarin1983a}, \citealt{Shandarin2010b}). Finally the DENMAX (DENsity MAXimum) halo finder assumes that the halos are the peaks of the density  fields and thus selects the particles concentrated in the vicinity of the density maxima \citep{Bertschinger1991}. One of the common features of these  techniques is that all three are based on density, in one form or another. And all of them depend on free parameters that are chosen chiefly on the `merits principle' \citep{Forero-Romero2009a} rather than on physics. Over the years all three kinds of the halo finders have been experiencing various modifications and improvements.  A few examples from a long list of these modifications may include:

(i) Improvised techniques of generation of the density field from the particle positions, and finding spherically bound over-densities (\citealt{Lacey1994}, \citealt{Jenkins2001}, \citealt{Evrard2002}, \citealt{Weinberg1997}, \citealt{Neyrinck2005}, \citealt{Knollmann2009a}, \citealt{Sutter2010a}, \citealt{Planelles2010} etc.) 

(ii) Adaptive methods controlling the linking length in methods using FOF (\citealt{Davis1985}, \citealt{vanKampen1995}, \citealt{Gottlober1999}, \citealt{Springel2001a}, \citealt{Habib2009a}, \citealt{Rasera2010} etc.)

(iii) Adaptive methods for searching the positions of density maxima (For example, \citealt{Klypin1999}.)

(iv) Generalization of FOF and DENMAX techniques to six-dimensional phase space, and many others (such as 6DFOF by \citealt{Diemand2006} and ROCKSTAR by \citealt{Behroozi2013a} use velocity-position space with parameters analogous to linking-length.)

(v) Computing hierarchical tree of clusters in the phase-space such as the Hierarchical Structure Finder \cite{Maciejewski2009c}, and the 6-D Hierarchical Over-density Tree \cite{Ascasibar2010}. 
(vi) Hybrid algorithms: frameworks such as the Hierarchical Bound-Tracing algorithm \cite{Han2012a} and SURV \cite{Giocoli2010a} identify haloes at multiple time steps from the simulation to find prospective sub-haloes. In addition, there are HOP methods by \cite{Eisenstein1998}, \cite{Tweed2009} and \cite{Skory2010}. 


A detailed comparisons of several halo/sub-halo finders is provided in \cite{Knebe2011a}, \cite{Knebe2013}, \cite{Onions2012}. In a nice summary discussing  these developments as well as describing a few new suggestions they  concluded that there was no general consensus for a precise definition of a halo or a sub-halo. Consequently, there were different estimates of number of haloes, halo mass functions, halo centre locations, boundaries and other parameters.

There are significant concerns with SO, DENMAX and FOF algorithms - both in terms of underlying mechanisms of halo formation and the parameters used in halo identification. SO is motivated by the analytical toy model of the collapse of a top-hat spherical density perturbation. Parameters of the virial radii $r_{vir}$ and virial mass $M_{vir}$ are determined by the regions with density $\rho_{vir} \geq \Delta_{vir} \times \rho_b$, where $\rho_b$ is the background density of the simulation box. $\Delta_{vir}$ is generally taken around $180$ or $200$, derived for an isolated spherical collapse model, and it varies for different cosmologies and redshift. The peaks in CDM models  not only aspherical, but their collapse is subject to tidal forces, mergers and presence of sub-structures - none of these complexities are weighed in the spherical collapse model.

For FOF, the free parameter of linking length is generally taken as $b = 0.2$ times the mean separation of particles at $z=0$. This inter-particle separation corresponds to $\Delta_{vir} \approx 180$ if the halo has an isothermal density profile, $\rho \propto r^{-2}$. Using percolation theory, \cite{More2011} argued that this corresponds to a rather wide range of over-densities depending on halo mass and density profiles. They found out that $b = 0.2$ corresponds to local over-density $\delta$ within the enclosed halo to be in the range of 250 to 600. Moreover, the resulting FOF-haloes need not have a compact geometry: often the haloes are irregularly shaped, which is unlikely if the halos are virialized.
Hence modern algorithms re-define the halo boundaries by excluding particles using post-processing techniques. In recent simulations with clear delineation of walls and filaments \cite{Angulo2013a}, $b=0.05$ was used for finding FOF-haloes since the traditional value of $b = 0.2$ corresponded to structures that percolate into the web structure.

Absence of dynamical traits in the FOF and SO algorithms are arguably more crucial. In phase-space, the halo collapse models show collisionless DM particles in oscillatory motions about a core, at successive foldings of the phase-space sheet. The velocity field within each oscillatory spiral is multi-valued in physical space. Incrementing {\it multistream} shells, separated by {\it caustic} surfaces sequentially trace the structures of the cosmic web - walls, filaments and the haloes. Majority of the mass accretion into the haloes along the filaments: from lower multistreams into higher. Thus the DM haloes are not independent of filaments around them, and the hierarchical layers of multistreams represents this portrait precisely. This picture of structure formation was initially theorized using ZA \cite{Zeldovich1970} and in context of caustics \cite{Arnold1982} as well as in the Adhesion Approximation (\citealt{Gurbatov1989}, \citealt{Kofman1992}). 
\cite{Shandarin1989} reviewed gravitational evolution of density perturbations in the context.  

It has been demonstrated that the multistream field in Eulerian space can be computed directly from the Lagrangian sub-manifold (\citealt{Shandarin2012} and \citealt{Abel2012}). About 90\% of the field is single-streaming voids, and the rest of the volume comprises of multistream walls, filaments and haloes. \cite{Ramachandra2015} found the multistream value of $n_{str} \approx 90$ corresponds to virial density $\Delta_{vir} = 200$. On the other hand, DM particles are identified by \citep{Falck2012} as belonging to haloes if they undergo flip-flop along 3 orthogonal axes. These analyses have opened up a new avenue in studies of halo formation, both qualitative and quantitative. Re-investigations of halo spins, physical radii of the halo, sub-structure in the light of streaming phenomena have shown that the halo structures and formations are more complicated than previously believed. \cite{Vogelsberger2011} investigated the distribution of streams in small haloes at various redshifts. They concluded that tracking caustics and streams is better than density, since density fields are noisy in the dense inner regions of haloes. In another study, \cite{More2015} argued that the `splashback radius' - distance from the halo core to the first caustic enumerated from outside - is a better physical indicator of DM halo boundary than the virial radius (also see the discussion on turn-around radius of bound objects by \citealt{Lee2016a}). \cite{Angulo2013a} also agree with the view that the locally overdense regions correspond better with the volumes within the first caustic than the virialized DM clumps. Recent toy model of anisotropic halo collapse by \cite{Neyrinck2016} considers intersecting multistream filaments forming spinning nodes. \cite{Ramachandra2017} showed that the virial surfaces of FOF haloes have varying number of streams, including single-streams. Study by \cite{Shandarin2016} delineated the rich sub-structure of haloes using another derivative parameter from the Lagrangian sub-manifold called the `flip-flop' defined on the Lagrangian space.


In this paper, we identify potential haloes by utilizing multistreaming as the governing dynamical phenomena. A review of the DM particle clustering in a one-dimensional dimensional universe is made in Section \ref{sec:HaloFormation}, and the concept of multistream field is extended to higher dimensions. The multistream field is computed on the cosmological simulations described in Section \ref{sec:simulation}. The halo identification framework in this field is described in Section \ref{sec:haloDetection}. This algorithm isolates convex regions of the multistream field using Hessian eigenvalues, each enclosing a local multistream maximum. Without employing any non-local thresholds that several halo finders generally use, these convex multistream regions are identified as potential halo sites. We also illustrate the significance of multistream refinement and softening scales in finding subhaloes. However, this paper does not focus on adaptive multi-scale analyses for substructure studies. A few properties of the multistream haloes are discussed in Section \ref{sub:HaloProperties}, and comparison of these haloes with AHF and FOF algorithms is done in Section \ref{sub:compareHalo}. We also discuss the spatial distribution of the dark matter haloes along the percolating web structure.

\section{Phase-space representation of gravitational clustering}
\label{sec:HaloFormation}

\begin{figure*} 
\centering\includegraphics[height=7cm]{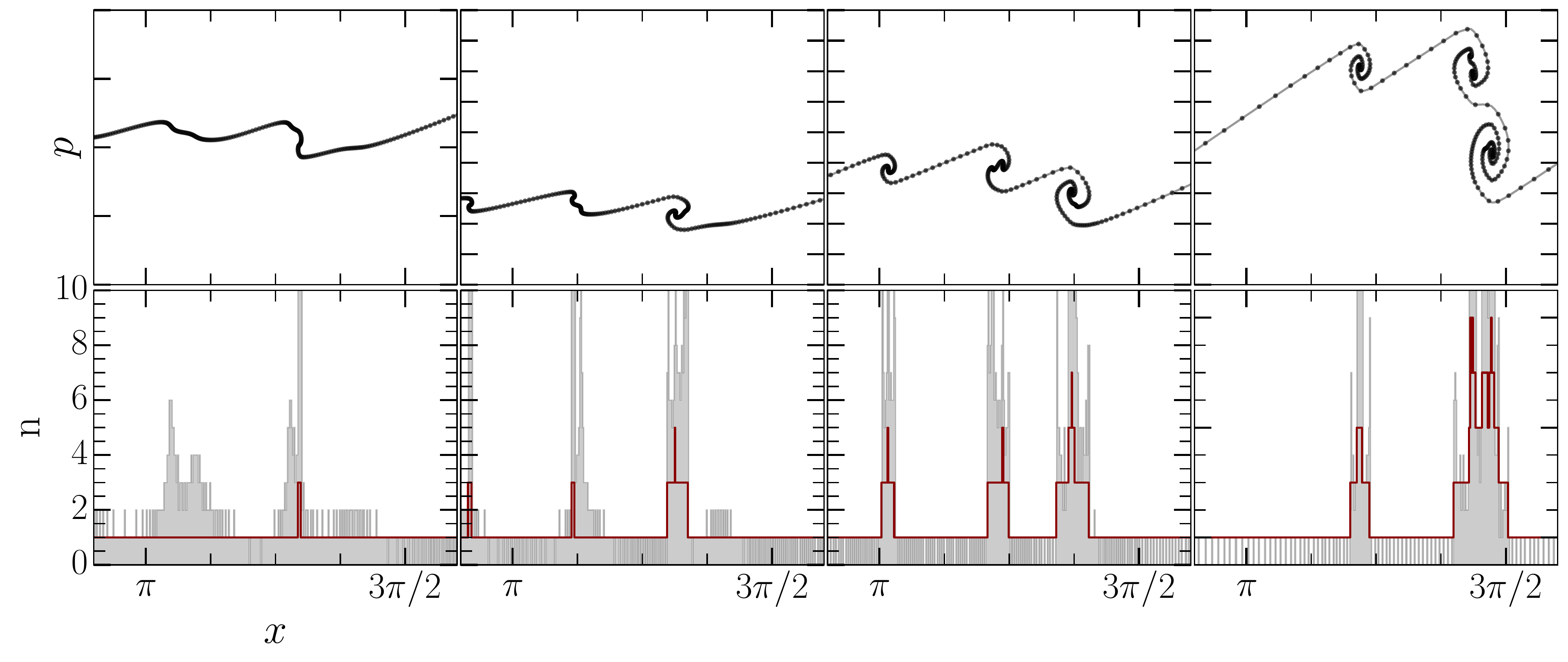} 
\caption{Dynamical collapse of dark matter in one-dimensional universe: Top panels show the $(\bmath{p}, \bmath{x})$ phase-space manifold of the dark matter sheet at redshifts $z_1$, $z_2$, $z_3$ and $z = 0$. Dots represent the dark matter particles. The momentum values are chosen at arbitrary scales. Bottom panels show the corresponding multistream field multistream field $n_{str}(\bmath{x}, z)$ (red) and density field $\rho(\bmath{x}, z)$ (gray). }
\label{fig:1d}
\end{figure*}

We begin with a simple illustration showing the formation of a few haloes in a one-dimensional simulation. Dark matter clustering in a (1+1)-dimensional phase-space $(\bmath{p}, \bmath{x})$ (where $p$ is the momentum and $x$ is the co-moving Eulerian coordinate) at four successive time steps is shown in the top panels of \autoref{fig:1d}. The lower panels show the corresponding multistream field (\citealt{Shandarin2012} and \citealt{Abel2012}) $n_{str}(\bmath{x}, z)$ (red) and density field $\rho(\bmath{x}, z)$ (gray). At $z_1$ (left-most panel), velocity is single-valued in Eulerian co-ordinates shown, except at a small three-stream region near $\bmath{x} = 5\pi/4$. This is the first instance of multistreaming in the region, which was previously had $n_{str} = 1$ throughout. The interface of $n_{str} = 1$ and $n_{str} = 3$ regions is also the location of the first caustic. On the other hand, the density calculated at a high resolution shows variations, even in the mono-streaming regions. The variations are especially more pronounced around  the caustic (near $\bmath{x} = 5\pi/4$).

The gravitational clustering is more evolved in the two center panels ($z_2$ and $z_3$) with three prominent phase-space spirals. The regions between the spirals have sparsely distributed dark matter particles, and have $n_{str} = 1$. Each spiral corresponds to a location of gravitational collapse with $n_{str} > 1$ region, and higher density. A few of these regions within three-streaming regions are elevated to $n_{str} = 5$. The corresponding density field is not only noisier, but also reaches very high values at the caustics. This is also a primary distinguishing feature between mass density fields and multistream fields: At the locations of caustic, the density (regardless of how it is calculated) is not smooth \cite{Vogelsberger2011}. Computational limitations on simulation resolutions and refinement of density calculations soften the fields, exceptionally at the zero volume measure regions of caustic surfaces.  On the other-hand, multistream values are increased by finite values at caustic surface locations - This property is preserved at higher simulation resolutions and any refinements of multistream field calculations - although $n_{str}$ may be resolved enough to have intermediate even-values. Multistream fields are also intrinsically discrete valued, which is not true with density fields. Discreteness of multistream fields is discussed in more detail in \cite{Ramachandra2017}. 

The right-most panel in \autoref{fig:1d} shows the final structure at $z=0$. Two large spirals have spatially merged. These collapse environments are naturally very complex, with an increased number of successive caustic formation and merging. 
The corresponding velocity streams also show a more complicated structure. Clearly, the multistream field has a saddle point that is not as low as $n_{str} = 1$. This poses a bigger problem in the context of most of halo detection algorithms, and we discuss this in Appendix \ref{appendix:Eigen}.

\subsection{Collapse in higher dimensions}

Extending the above results of one-dimensional collapse into higher dimension is vital, primarily in the context of halo formation. The individual spiral collapses in the one-dimension happen at a small region (left-most panel in \autoref{fig:1d}), and the region grows by volume, whilst increasing the spiral twists within. This is in contrast with the theoretical top-hat spherical model of halo formation when the shell crossing would not happen until the final moment of the collapse of the entire halo into a point-like singularity. Thus all shell crossings happen at a single point
and at a single instant of time. The collapse of an isolated, spherically symmetric density peak is a very exceptional case, because every spherical shell feels only the force due to interior mass until it collapses into the caustic region. The collapse of the real peak proceeds in the field generated by the mass distribution - in both the mass within the forming halo, and the mass outside the halo. 

The collapse of a uniform ellipsoid also results in a simultaneous collapse of the entire ellipsoid
however this time not into a point but into a two-dimensional ellipse (\citealt{Lin1965}, \citealt{Icke1975}, \citealt{Eisenstein1995}).
Another customarily used spherical model of halo formation by \cite{Fillmore1984} and \cite{Bertschinger1985} does not consider
the initial collapse at all. Instead it assumes self-similar initial conditions and the halo at advanced stage with formally infinite number of spherical caustic shells.

The `core' in a collision-less dark matter collapse (in \autoref{fig:1d}) is a region where a multistream region is first formed due to caustic formation. This is conceptually similar to a shell crossing. However, there are successive caustic formations that follow the first shell crossing, and they are not limited to the halo cores. Each caustic increases the multistream value within by a finite number. The cores of the multistream haloes obviously have the local maxima of velocity streams in Eulerian coordinates. On the contrary, mass densities have infinite values at the caustics surfaces, including the core. Discontinuities in densities at these regions of sharp multistream transitions are clearly seen if the mass and spatial resolutions were sufficiently high( see two-dimensional simulations by \citealt{Melott1989} as well as in three-dimensional simulations by \citealt{Hahn2013}, \citealt{Angulo2016}, \citealt{Hahn2016a} etc.). 

\begin{figure} 
\centering\includegraphics[width=8cm]{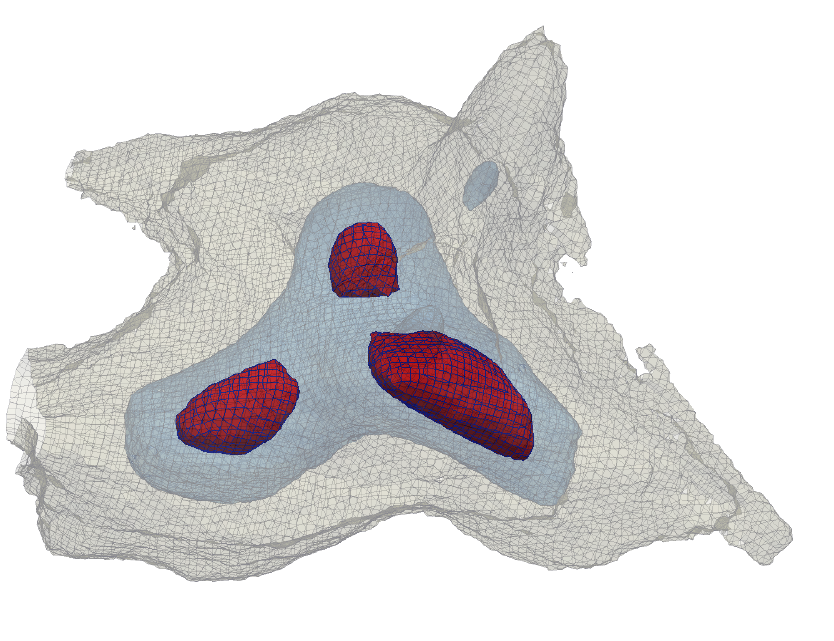} 
\caption{Multistream field contours: The multistream field is calculated at 8 times the native resolution. Each inner convex blobs (red) surround local multistream maxima inside. Surrounding outer shell(blue) is not convex throughout the surface, and the outermost gray multistream surface displays a filamental geometry.}
\label{fig:full}
\end{figure}

In three-dimensional simulations, the Lagrangian sub-manifold twists in complicated ways in a six-dimensional phase space. This is due to complexities involving caustic formations in higher dimensions, which is true even in the ZA, see \citealt{Arnold1982} and \citealt{Hidding2014} for detailed analyses of caustic formation. The resulting intricate geometrical structures can be characterized by the $n_{str}$ field. Nearly  $90\%$ of the volume in N-body simulations are single-streamed voids at $z=0$ (\citealt{Shandarin2012}, also see \citealt{Falck2015} for a percolation analysis of single-streaming voids). From the visualizations in \cite{Ramachandra2015} and percolation analysis of \cite{Ramachandra2017}, we also know that the $n_{str} = 3$ regions mostly form connected wall-like structures, unlike the isolated patches as seen in one-dimensional simulations of \autoref{fig:1d}. The structures become predominantly filamentary at higher thresholds of $ n_{str} \gtrsim 17$. Subsequently, the regions around the multistream maxima have isolated closed surfaces (for example, in \autoref{fig:full}), which may be identified as halo locations. 

Caustic formations and mass accretion are also seen to occur more along the higher streams, which makes the haloes non-spherical, with the alignment generally determined by a complicated interplay  of the intensities of the streams from neighboring filamentary structures. Number of streams corresponding to the dark matter halo also has a local environment dependence. The three small haloes in \autoref{fig:full}, where the number of streams are higher than the neighbouring filaments, are aligned along three intersecting filaments. Halo environment studied in \cite{Ramachandra2015} show similar hierarchical variation in $n_{str}$ values. The halo environments are thus very complicated, and empirical thresholds (on multistream or density fields) may not account for all the haloes uniformly. Hence we use a local geometrical method to identify potential haloes in multistream fields.

The first non-linear structures in the web have $n_{str} = 3$. By visual inspection, these regions generally form a fabric-like open structures that resemble walls. The surface contours of higher $n_{str}$ are embedded within the walls. \autoref{fig:full} shows a filamentary structure of the web at $n_{str} \gtrsim 17$. The figure also shows regions around local maxima of the multistream field, which are generally located at the intersections of filaments.

\section{Simulations and tools}
\label{sec:simulation}

The emphasis of this paper is to demonstrate the use of multistream field in identifying potential dark matter haloes, and not a full statistical analysis of halo properties. For this purpose, we have run simulations at two different mass resolutions (number of particles $N_p = 128^3$ and $256^3$, and respective mass of particles, $m_p = 3.65 \times 10^{10} h^{-1} M_{\sun}$ and $ 4.57 \times 10^{9} h^{-1} M_{\sun}$), with the same periodic side length $L = 100 h^{-1} Mpc$. The gravitational softening length $\epsilon = 20 h^{-1} kpc$ and $10 h^{-1} kpc$ for low and high resolution simulation respectively. The initial conditions are generated by {MUSIC} \citep{Hahn2011a} with the transfer function from \cite{Eisenstein1998a} at a redshift of $z_{ini}= 80$. The $\Lambda$CDM cosmological simulation is run using {GADGET-2} (\citealt{Springel2005} and \citealt{Springel2001}) is similar to the ones used in \cite{Ramachandra2017}. The cosmological parameters used in the simulation are $\Omega_{m}= 0.276$, $\Omega_{\Lambda}= 0.724$, the Hubble parameter, $h = 0.703$, the power spectrum normalization, $\sigma_8 = 0.811$ and the spectral index $n_s= 0.961$.

Multistream field $n_{str}(\mathbf{x})$ is calculated on the GADGET-2 snapshots at $z=0$ using the tessellation scheme by \cite{Shandarin2012}. The multistream field can be computed at the native resolution of the Lagrangian grid of the simulation, i.e., at refinement factor of $l_l/l_d = 1$ (where $l_l$ is the inter-particle separation in Lagrangian grid and $l_d$ is the side length resolution of diagnostic grid). Arbitrarily high refinement factors can be utilized in computing multistream fields as well, for example  $l_l/l_d = 8$ for the halo multistream environment shown in \autoref{fig:full}. For analysis of full simulation boxes, we restrict $l_l/l_d$ to $1$ and $2$. 

Two halo finders are also used to identify potential haloes with 20 or more particles at $z= 0$: a classic Friends-of-Friends method (FOF-\citealt{Davis1985}) using a popular linking length, $ b= 0.2$ (e.g. \citealt{Frenk1988} and \citealt{Lacey1994}) and the Adaptive Mesh Investigations of Galaxy Assembly (AMIGA halo finder or AHF-\citealt{Knollmann2009a}, \citealt{Gill2004a}). Halo catalogue from these halo finders are used to compare with our implementation of halo detection in the multistream field. The haloes candidates from AHF and FOF algorithms are hereafter referred to as AHF-haloes and FOF-haloes respectively.

\section{Haloes in the multistream field}
\label{sec:haloDetection}

We intend to identify haloes in the $n_{str}(\bmath{x})$ field instead of using just the position coordinate data. While the eigenvalue analysis itself is done at a chosen time, the multistream field inherently has data from six-dimensional Lagrangian space $(\bmath{q}, \bmath{x})$ that contains the full dynamical information,similar to the phase-space sheet albeit in a different form. Dynamical history that is embedded in the multistream field is crucial in understanding the strength of gravitational binding of the particles. 
A physically motivated distinction between void and gravitationally collapsed regions -- voids are the regions with a single stream -- is a unique feature of multistream analysis (\citealt{Shandarin2012} and \citealt{Ramachandra2017}). Thus the haloes detected from local maxima of the $n_{str}$ field can be ensured to be away from the mono-streaming voids. Methods based on linking-length or density fields may not be able to ensure that  all the particles in haloes are away from voids (as shown for FOF haloes in \citealt{Ramachandra2017}) .

Numerical analyses of scalar fields generally depend on resolution as opposed to particle coordinates analysis tools like FOF. The multistream field conveniently has an advantage of being less noisy than mass density (\citealt{Shandarin2012}, also see the Appendix in \citealt{Ramachandra2017} ). 

\subsection{Hessian of multistream fields}

Hessian matrix $\mathbfss{H}(f)$ of a scalar field $f$ involves local second-order variations in three orthogonal directions. Each element of the Hessian matrix $\mathbfss{H}_{ij}(f)$ (where $i$ and $j$ can be any of $x$, $y$ or $z$ directions) is given by \autoref{eq:Hess}. 

\begin{equation}
\label{eq:Hess}
 \mathbfss{H}_{ij}(f) = \frac{\partial^2 f}{\partial x_i \partial x_j}
\end{equation}

By choosing a $f = -n_{str}(x)$ (smoothened if necessary), local multistream variations can be diagnosed. The Hessian matrices at each point on the configuration space are always symmetric matrices, resulting in real eigenvalues. The Hessian eigenvalues in multistream fields differ from that in density, gravitational potential or velocity shear tensor. We refer the readers to \cite{Ramachandra2017} for an extensive analysis on multistream Hessians and their geometrical significance. Some of the salient features of Hessian eigenvalues of multistream field are as follows: 

(i) Every element of Hessian matrices $\mathbfss{H}(-n_{str})$, and consequently the eigenvalues $\lambda_i$'s are zero in single-stream voids. Even if the multistream field is a smoothed, the eigenvalues peak at zero. This property is unique to multistream fields. Eigenvalues of Hessians of density \cite{Aragon-Calvo2007}, velocity shear tensor \cite{Libeskind2013} do not peak at zero, and the eigenvalues of deformation tensor are negative in voids as a result of continuity equation (shown in Zel’dovich formalism as well).

(ii) The eigenvalues of these Hessian matrices are always real, and depending on if their values are positive or negative, one may infer local geometrical features in the field. For our choice of $-n_{str}(x)$ as the domain of Hessian, at least in principle, the conditions for geometric criteria are: $\lambda_1 > 0 > \lambda_2 \geq \lambda_3$ for locally flat regions, $\lambda_1 \geq \lambda_2 > 0 > \lambda_3$ for locally tubular structures and $\lambda_1 \geq \lambda_2 \geq \lambda_3 > 0$ for clumped blobs. 

(iii) Convex neighbourhoods around local maxima of the multistream field are isolated by the positive definite Hessian matrices. 

(iv) The resulting Hessian eigenvalues characterize the geometry in a four-dimensional
hyper-space of $(-n_{str}, x, y, z)$. The boundary of a region with $\lambda_1 \geq \lambda_2 \geq \lambda_3 > 0$ is a closed convex contour in this hyper-space, and thus it’s projection on to the three-dimensional Lagrangian space is also closed and convex.

Of the three geometries that are characterised by the eigenvalue conditions, we investigate the convexity of multistream contours in the context of halo finding in the section below.

\subsection{Halo finder algorithm}
\label{subsec:technique}

{Our goal is to isolate the locations of convex geometries in the multistream flow field. Prospective regions of the halo locations in the web structure are selected by positive definite condition of the Hessian $\mathbfss{H}(-n_{str})$: $\lambda_1 > 0$, $\lambda_2 > 0$ and  $\lambda_3 > 0$, or simply the smallest of the eigenvalues, $\lambda_3 > 0$. We also filter out the regions if the multistream values inside do not suggest gravitational collapse into haloes. The sequence of our halo detection framework is listed below:

\begin{enumerate}
\item The multistream flow field is calculated on a diagnostic grid. The number of tetrahedra that encompass each vertex in the grid gives the $n_{str}$ field. Top left panel of \autoref{fig:labelsfilter} shows the multistream web structure in a slice of the simulation with $n_{str} > 1$ in gray and $n_{str} \geq 7$ in blue.  

\item The discrete multistream flow field is smoothed in order to reduce numerical noise. We have used Gaussian kernel for smoothing in our analysis. Effect of smoothing scales in the halo identification is discussed in Section \ref{sub:Smooth}. 

\item Second order variations of the smoothed $-n_{str}(\bmath{x})$ is computed at each point in the field. This gives symmetric Hessian matrices for this field whose eigenvalues are real. Ordered eigenvalues of the Hessian, $ \lambda_1 \geq  \lambda_2 \geq \lambda_3$ are calculated. The $\lambda_3$ field is shown in the top right panel of \autoref{fig:labelsfilter}. 

\item Using segmentation techniques, each region with $ \lambda_3 $ strictly greater than $0$ within $n_{str} \geq 3$ regions of multistream field are isolated and labelled. This condition for each halo candidate guarantees that it is in the region where at least one gravitational collapse happened within the halo boundary. Mass particles belonging to these regions are shown shown as dark spots in in the top right panel of \autoref{fig:labelsfilter}. 

\item The multistream field has a range of values within the isolated sites. We impose constraints on the isolated regions to rule out the labels with low multistream values. The local maxima of $n_{str}$ inside each halo must be at least 7. By this restriction, it is ensured that the halo sites with three Lagrangian sub-manifold foldings are selected. Bottom left panel of \autoref{fig:labelsfilter} shows patches that are discarded in red. The resulting $\lambda_3$-haloes are shown in the bottom right.

\item In our comparisons with other halo finders in Section \ref{sub:compareHalo}, we also used an additional constraint on the minimum number of mass particles in the haloes to be 20 - which is generally used as a criteria in several halo finders. 

\end{enumerate}

\begin{figure}
\begin{minipage}[t]{0.99\linewidth}
 \centering\includegraphics[height=8.5cm]{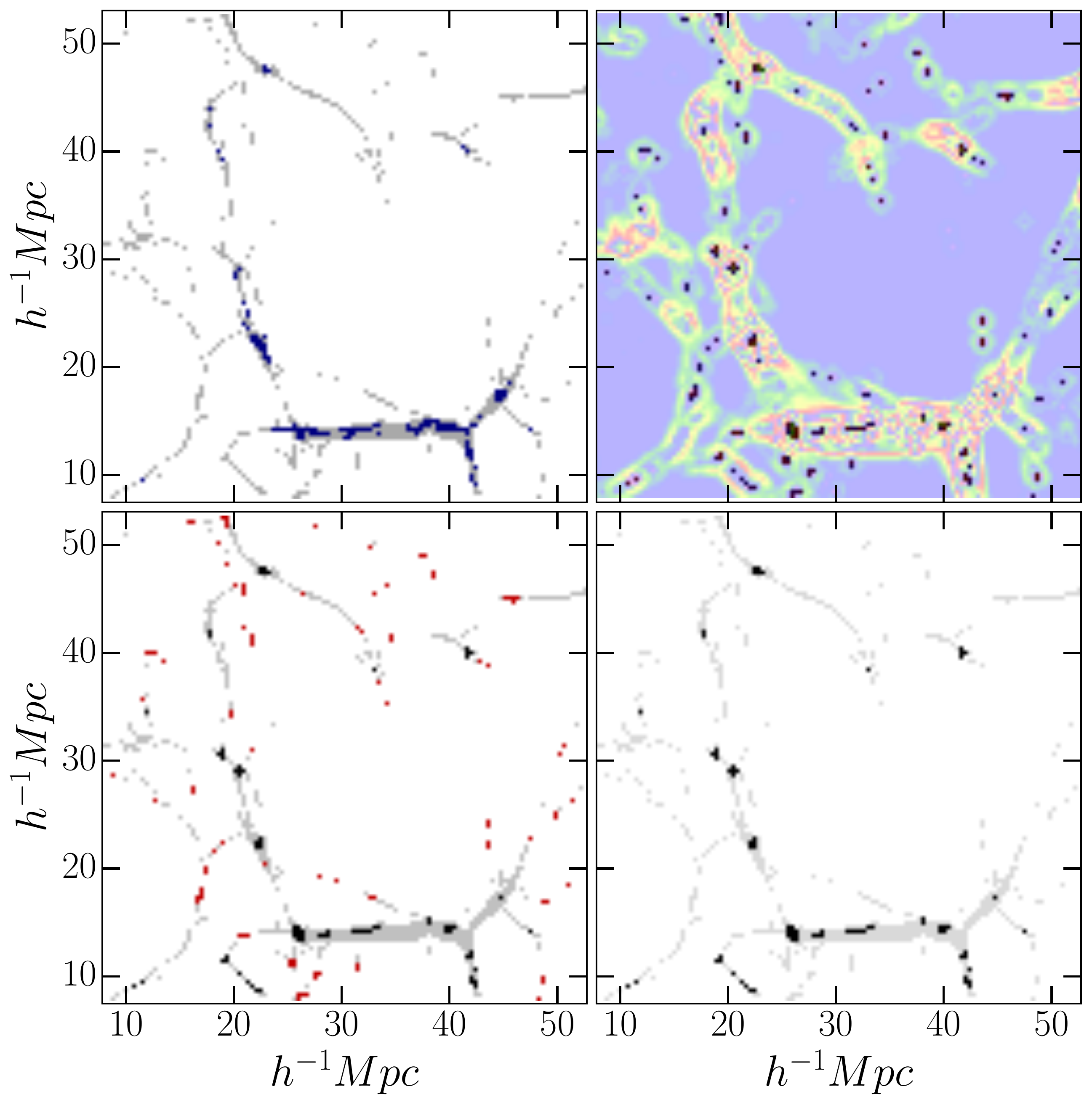} 
\end{minipage}\hfill
\caption{ Detection of potential halo candidates in the multistream field: algorithm of segmentation and filtering are illustrated in a smaller slice of $40 h^{-1} Mpc \times 40 h^{-1} Mpc$ slice of the simulation box. Top left figure shows the multistream field of the slice. Voids (white) are the regions with $n_{str} =1$, rest are non-void structures. Blue patches within the structure (gray) are the regions with gravitational collapses in more than one direction, i.e., $n_{str} \geq 7$. Top Right figure shows the smallest eigenvalue $\lambda_3$ field. The value of $\lambda_3$ is close to 0 in most of the regions (yellow), including the voids. Regions with  $\lambda_3 > 0$ and $n_{str} > 1$, are isolated (black spots) using image segmentation techniques. Bottom left panel shows the filtering scheme: the red patches do not have maxima of $n_{str} \geq 7$ in the regions, hence are filtered out. The remaining potential halo regions with more than 20 particles are shown in the bottom right panel.}
\label{fig:labelsfilter}
\end{figure}

For the illustration halo detection framework in this section, we have calculated the number-of-streams at refinement factor of $2$ and smoothing scale of $0.39 h^{-1} Mpc$ (equal to the grid length of the multistream field) for the simulation box of $128^3$ particles and size $L = 100 h^{-1} Mpc$ . Hessian matrices and eigenvalues are calculated on the same diagnosis grid. Results of the halo detection scheme for simulation box of higher mass resolution, and different smoothing factors are discussed in Sections \ref{sub:compareHalo} and \ref{sub:Smooth}. Hereafter we refer to the potential dark matter haloes detected from the Hessian analysis of the multistream field as $\lambda_3$-haloes for brevity.

Applying the above scheme on the simulation with side length of $100 h^{-1}$ Mpc and $128^3$ particles (with cosmological parameters mentioned in Section \ref{sec:simulation}), we detected approximately 50000 regions satisfying $\lambda_3 > 0$ within the non-void in the multistream field of refinement factor $l_l/l_d = 2$ and smoothing scale of grid length, i.e, $0.39 h^{-1} Mpc$. We filtered out the segments with local maxima of $n_{str} < 7$, and around 14000 regions remained as prospective haloes. Majority of these regions have less than 20 particles, which are excluded in the halo catalogues. On the whole, our algorithm detected about 4500 haloes with more than 20 particles in the entire simulation box. We have not applied virialisation to define the halo boundaries. A more detailed study of halo edges, and comparison with that of FOF-haloes and AHF-haloes is done in Section \ref{sub:HaloProperties}. Here we concentrate on the three vital factors in our framework: local geometrical indicators $\lambda_i$'s, the softening scale of the field and multistream thresholds.

The maximum values of $\lambda_1$, $\lambda_2$ and $\lambda_3$ in each of the haloes have peaks away from 0 as shown in \autoref{fig:maxL3}. The median values of $max(\lambda_1)$ and $max(\lambda_2)$ are in the range of 1-10 (\autoref{tab:maxL3}), in-spite of the threshold for $\lambda_3$ being barely positive, by definition. Hence the interior of the potential halo segments is quite convex, with a local maxima inside. In some haloes, the local maxima of eigenvalue are in the order of thousands, as tabulated in \autoref{tab:maxL3}.

\begin{figure}
\begin{minipage}[t]{.99\linewidth}
 \centering\includegraphics[width=8.cm]{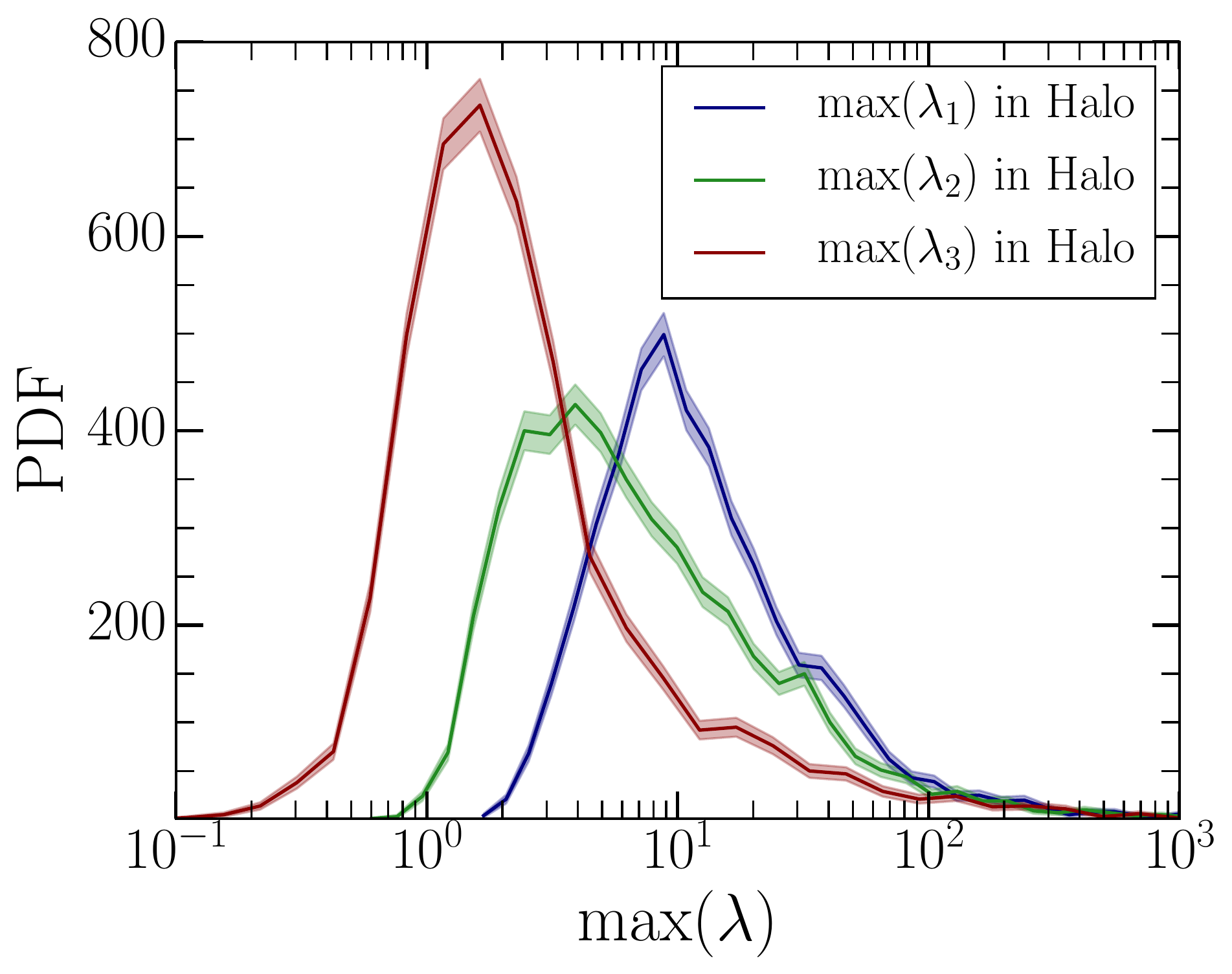} 
\end{minipage}\hfill
\caption{PDF of highest $\lambda_1$, $\lambda_2$ and $\lambda_3$ values in each of 4492 haloes detected by out algorithm. The peaks of the PDF are in the range 1-10. Shaded regions represent 1$\sigma$ error. }
\label{fig:maxL3}
\end{figure}

\begin{table}
  \caption{Statistics of the Hessian eigenvalues in the halo candidates}
\begin{tabular}{|l|r|r|r|}
\hline
Statistics                &  $\lambda_1$ &  $\lambda_2$ &  $\lambda_3$\\ \hline
Minimum   & $ 1.5 $ & $ 0.5 $ & $ 1.3 \times 10^{-2} $  \\ \hline
Maximum  & $1.7 \times 10^3$  & $1.5 \times 10^3$ & $1.1 \times 10^3$     \\ \hline
Median     & $10.5$  & $5.5$ & $1.9$  \\ \hline
\end{tabular}
\label{tab:maxL3}
\end{table}

With this algorithm, we obtain prospective dark matter haloes - regions with a local maximum of the multistream field in the interior of their closed convex surfaces. The haloes are detected without using density fields or linking lengths between particles. The parameters in the algorithm are entirely based on features of the multistream field and local geometry using Hessian matrices.

\subsection{Effect of smoothing}
\label{sub:Smooth}

In order to reduce noise, the field is smoothed for our analysis using a Gaussian filter. The effect of smoothing scale on the distribution of the eigenvalue $\lambda_3$ in the simulation of $128^3$ particles is shown in \autoref{fig:Eval3Smooth}. Effect of softening on the multistream fields does not alter the distribution of multistream distribution significantly (Seen in Figure 10 of \citealt{Ramachandra2017}).
However, the second order variation (and consequently the Hessian eigenvalues) is significantly changed due to the softening of the edges of structures. PDF of $\lambda_3$ at multistream smoothing scale of the half the side length of diagnostic grid , $0.5 \times l_d = 0.20 h^{-1} \text{ Mpc}$ is noisier than in the scales of $l_d$ and $2 \times l_d$. However, at every scale, the PDF peaks at $0$. The volume fraction of regions with $\lambda_3 > 0$ (i.e. with positive curvature) is $2.4\%$, $2.3\%$ and $2.5\%$ for scales $0.20 h^{-1} \text{ Mpc}$, $0.39 h^{-1} \text{ Mpc}$, $0.78 h^{-1} \text{ Mpc}$ respectively. For the detection of haloes in Section \ref{sec:haloDetection}, we only look at these regions.

\begin{figure}
\begin{minipage}[t]{.99\linewidth}
 \centering\includegraphics[width=8.cm]{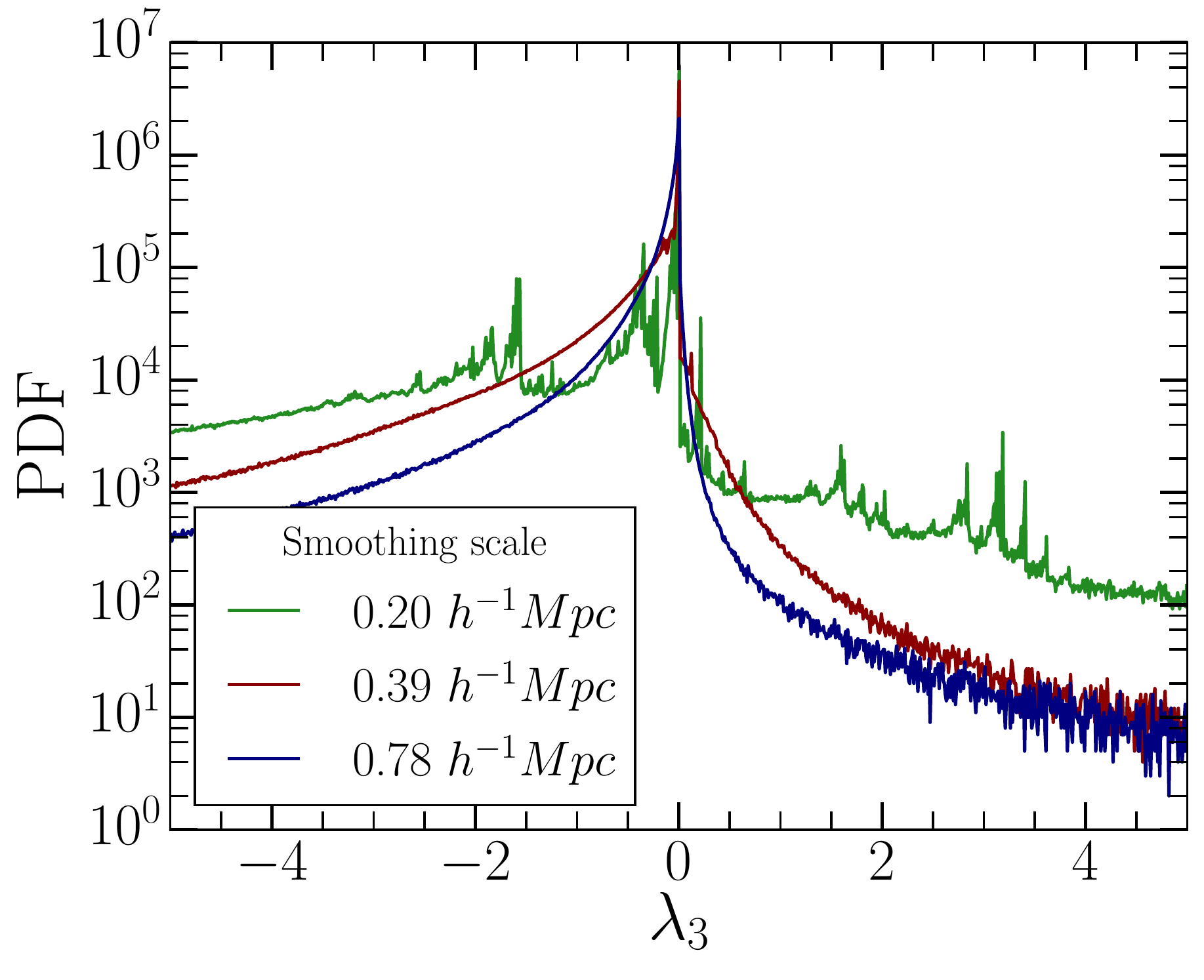} 
\end{minipage}\hfill
\caption{The distribution of $\lambda_3$ in the simulation box of $128^3$ particles and multistream field of refinement factor $l_l/l_d = 2$. Three smoothing scales are shown. }
\label{fig:Eval3Smooth}
\end{figure}

In addition to reducing the numerical noise, smoothing of the multistream field also results in softening of the sub-structures \cite{Ramachandra2017}. Since our framework of detecting haloes isolates the multistream regions with local maxima, the closed curvatures are resolved separately. The halo or a sub-halo regions, that enclose the local maxima of the $n_{str}$ field, vary with the smoothing scale of the multistream field. By increasing smoothing of the multistream field, the number of haloes are reduced as shown in \autoref{tab:HaloesSmooth}. In the simulation with $256^3$ particles, $27929$ $\lambda_3$-haloes are detected at smoothing scale equal to the diagnostic grid length, $l_d = 0.20 h^{-1} \text{Mpc}$. The number of haloes decreases to $18221$ and $7897$ at softening scales of two- and four times the grid lengths respectively. 

\begin{table}
  \caption{Number of $\lambda_3$-haloes identified at smoothing of $n_{str}(\bmath{x})$ at different scales.}
\begin{tabular}{|l|r|r|r|}
\hline
$N_p$  &  $0.20h^{-1} \text{Mpc}$ & $0.39 h^{-1} \text{Mpc}$ & $0.78 h^{-1} \text{Mpc}$ \\ \hline
$128^3$   & $5181$  &  $4492$ & $2923$ \\ \hline
$256^3$   & $27929$  & $18221$ & $7897$ \\ \hline

\end{tabular}
\label{tab:HaloesSmooth}
\end{table}

Moreover, since the spatial resolution is higher at the low softening, more small haloes are detected, as shown in lower mass regime of halo mass functions in \autoref{fig:hmfSmooth}. The tail of halo mass functions reveal that large haloes are more massive for higher softening scales. For instance, the largest haloes for the same simulation with multistream softening length of $0.20 h^{-1} \text{Mpc}$, $0.39 h^{-1} \text{Mpc}$ and $0.78 h^{-1} \text{Mpc}$ have $30650$, $38333$ and $56257$ particles respectively. 

\begin{figure}
\begin{minipage}[t]{.99\linewidth}
 \centering\includegraphics[width=8.cm]{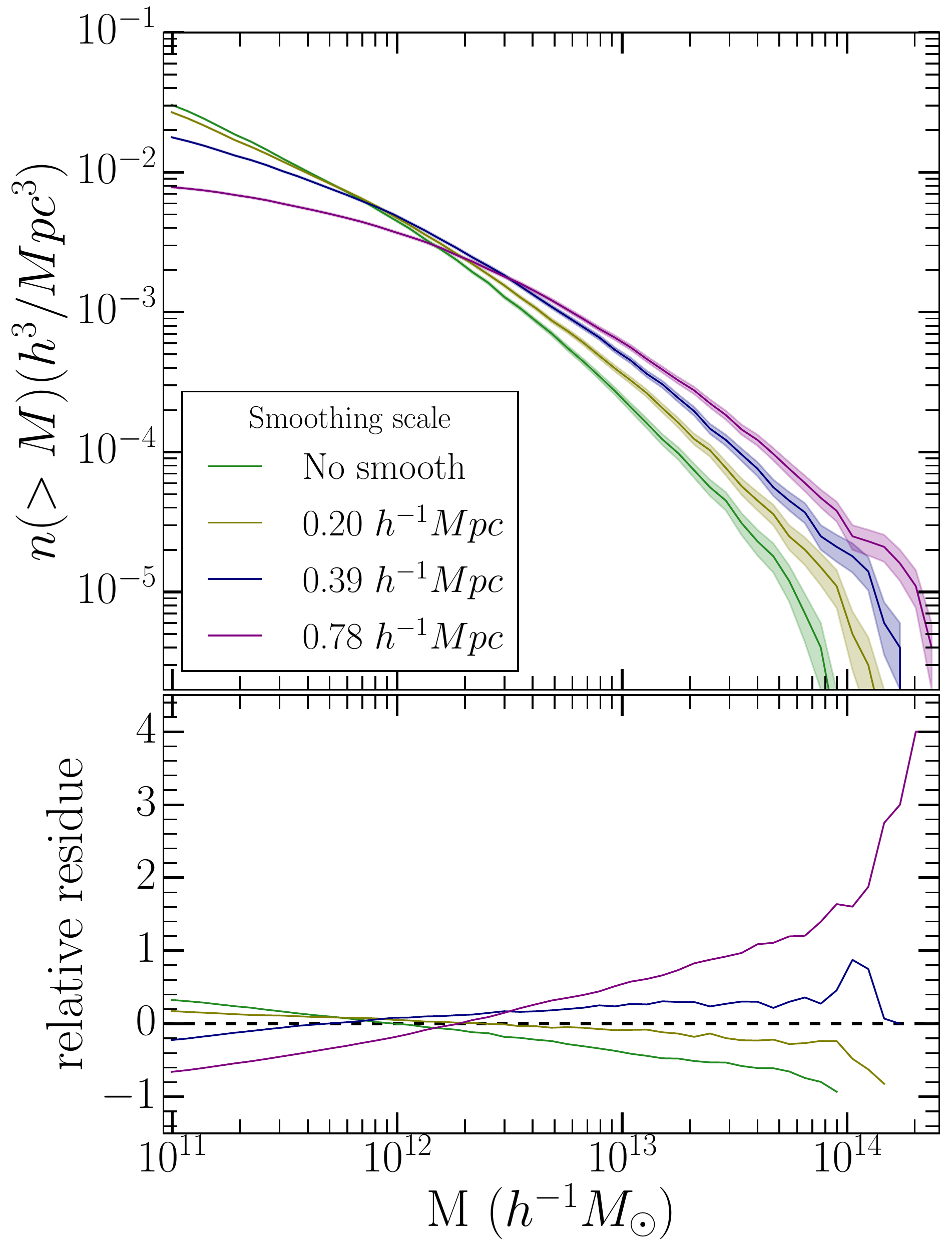} 
\end{minipage}\hfill
\caption{Top panel shows Halo Mass functions of the potential $\lambda_3$-haloes in the multistream field refinement factor $l_l/l_d = 2$ with various smoothing scales. Simulation box has $256^3$ particles. Lower panel shows the deviation of the each halo mass function with respect to their average.}
\label{fig:hmfSmooth}
\end{figure}

The sub-halo finder methods (see \citealt{Onions2012} and references therein) identify substructures within a large host halo. The sub-haloes are resolved individually as $\lambda_3$-haloes at different scales from our algorithm if the local maxima of the smoothed multistream field is enclosed within the boundary.

\subsection{Effect of multistream thresholds}
\label{sub:multiThresholds}

Environmental dependence of the haloes results in various multistream values for the halo core. Theoretical toy models of halo formation, such as the tetrahedral collapse model \cite{Neyrinck2016} describes a three-dimensional halo with four filaments accreting mass into it, has 15 stream crossings.  
\cite{Ramachandra2015} have previously showed that a high threshold of $n_{str} \geq 90$ is equivalent of virial density of $\rho_{vir} = 200$, and filters most of the large haloes above $10^{13} M_{\sun}$. 

The algorithm used for detecting multistream haloes initially detects all the closed regions in the multistream ($n_{str} > 1$) regions of the cosmological simulation. In order to exclude some of the obvious non-halo sites, we impose a lower threshold of $n_{str} \geq 7$ on the multistream maximum (these regions were also seen as parts of walls or filaments in \citealt{Ramachandra2015} ), so that all the sites with three or more foldings in the Lagrangian sub-manifold are chosen. Combining this with the conditions on local eigenvalues, number of particles in haloes etc, we got a pretty good correspondence with other halo finders as demonstrated in \autoref{sub:compareHalo} . 

Although this condition is by no means strict, it is necessary to check the validity of the assumption. \autoref{fig:hmfEth} shows the halo mass functions for the haloes detected with changing thresholds on the multistream values of the halo cores. The figure demonstrates that increasing the cut-off from $n_{str} \geq 3$ to $n_{str} \geq 25$ systematically excludes small mass
haloes while the number of haloes with $M  \gtrsim 2\times10^{12} M_{\bigodot}$  remains the same.}

\begin{figure}
\begin{minipage}[t]{.99\linewidth}
 \centering\includegraphics[width=8.cm]{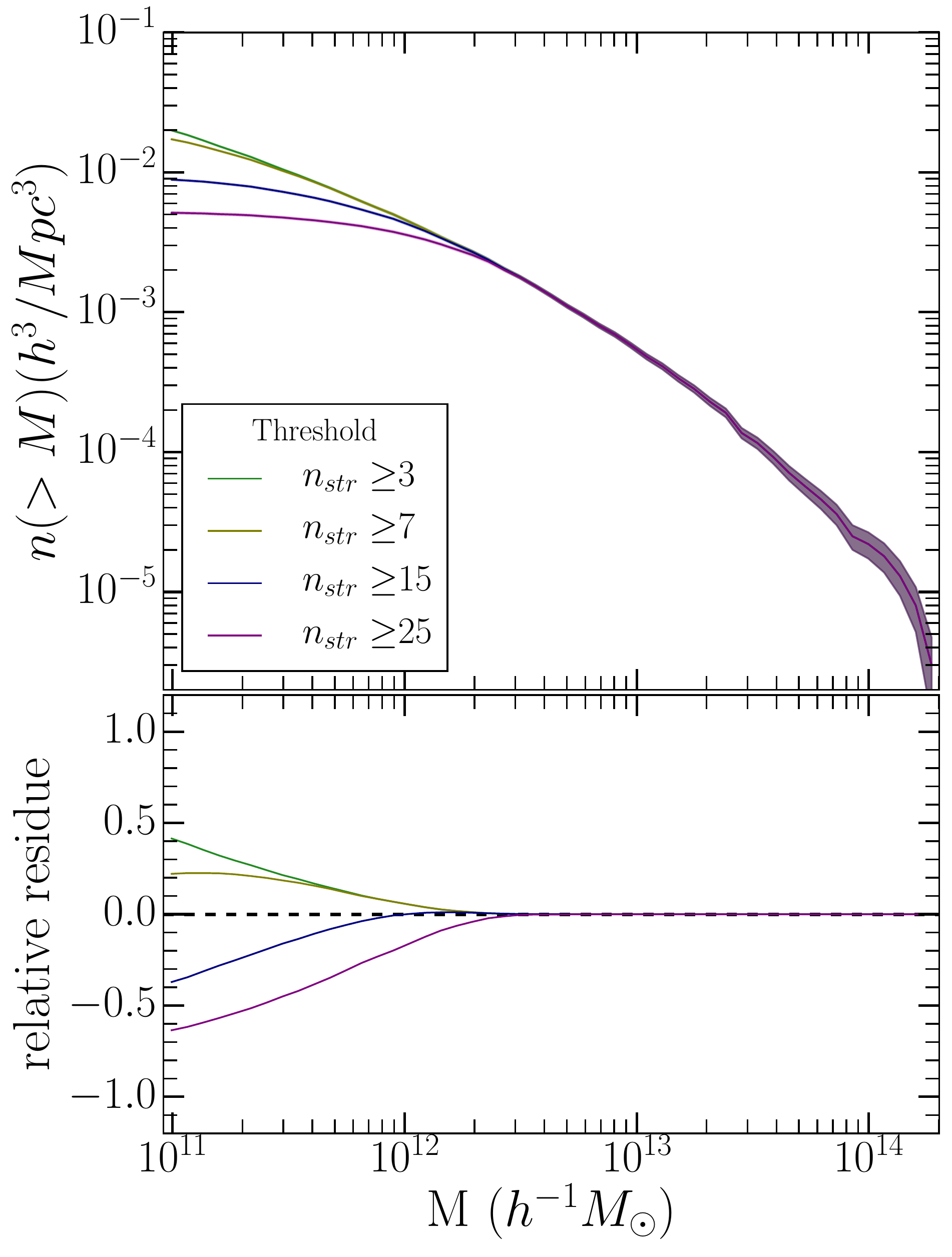} 
\end{minipage}\hfill
\caption{Halo Mass functions of the potential $\lambda_3$-haloes in the multistream field refinement factor $l_l/l_d = 2$ (in Simulation box with $N_p = 256^3$) with various thresholds on local maxima of $n_{str}$ within the halo. Lower panel shows the deviation of the each halo mass function with respect to their average.}
\label{fig:hmfEth}
\end{figure}

\section{Halo properties}
\label{sub:HaloProperties}

Multistream environment of haloes can be very diverse. \cite{Ramachandra2015} demonstrated that a majority of the FOF-haloes are in contact with the single-streaming voids. Illustration in \cite{Ramachandra2017} also shows that a large number of FOF-haloes have more than 10 per cent void on the spherical surfaces with virial radii. The $\lambda_3$ haloes are significantly different: none of the $\lambda_3$-haloes are in contact with the regions where gravitational collapse has not occurred. This is guaranteed by the lower bound of $n_{str} = 3$ on all potential halo candidates. Condition on the multistream field within the potential halo sites also ensures that there are collapses along more than one direction, which corresponds to $n_{str} = 7$. Hence by definition, for any multistream halo $H_i$, highest and the lowest multistream value are $n_{str}^{high}(H_i) \geq 7$ and $n_{str}^{low}(H_i) \geq 3$ respectively.

\begin{figure}
\begin{minipage}[t]{.99\linewidth}
 \centering\includegraphics[width=8.cm]{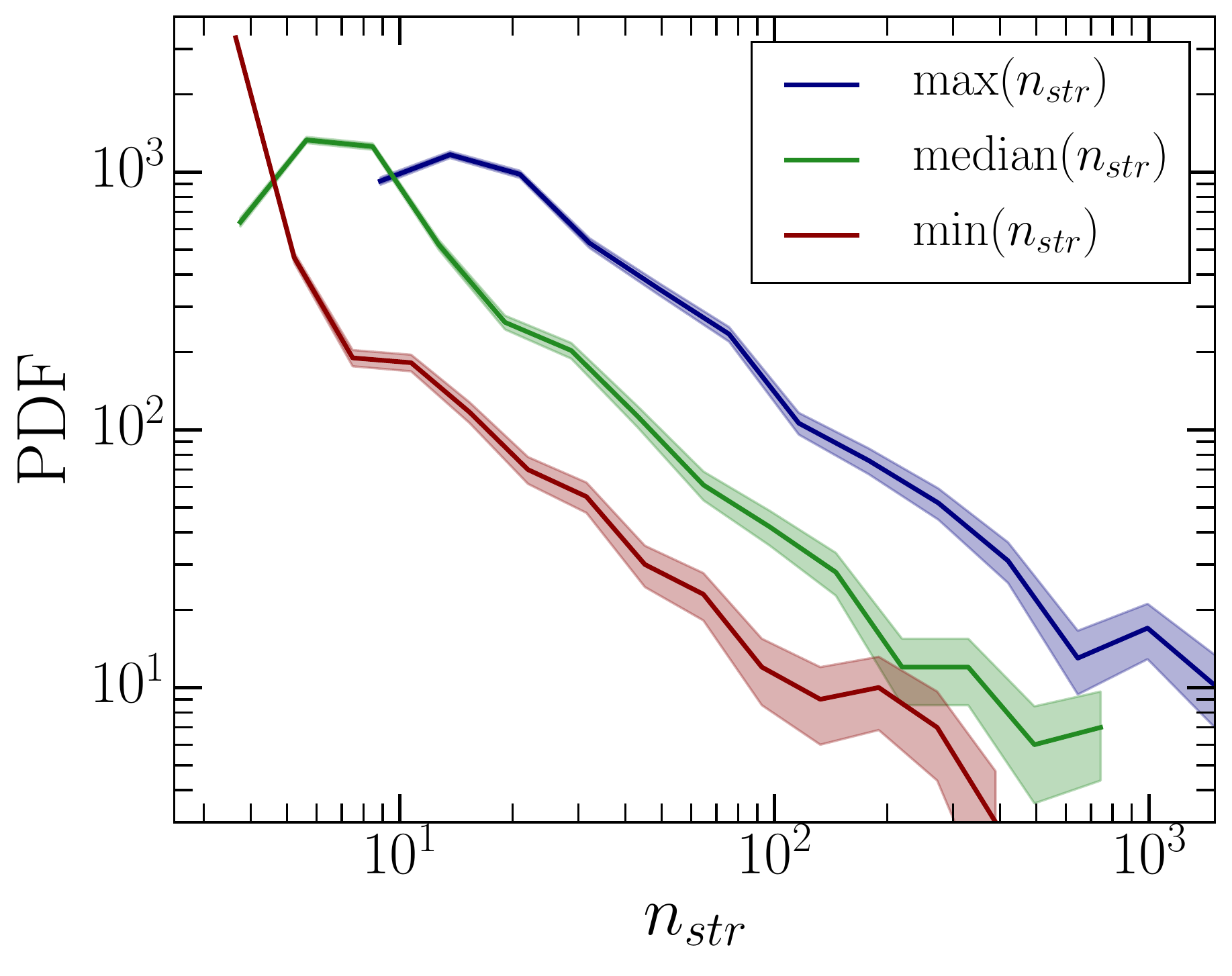} 
\end{minipage}\hfill
\caption{Maximum, minimum and median of $n_{str}$ in each of 4492 halo candidates. The closed contours of haloes encompass a wide range of multistream values. None of the haloes are in contact with the void region, since lowest value of min$(n_{str})$ is 3. Shaded regions are the 1$\sigma$ absolute errors in the number of $\lambda_3$-haloes. }
\label{fig:minmaxstr}
\end{figure}

The potential haloes $H_i$s selected by eigenvalue condition $\lambda_3 > 0$ have a local maxima of $n_{str}^{high}(H_i)$ inside their boundaries. For a large number of these $\lambda_3$-halo candidates, the maximum $n_{str}^{high}$ is higher than the bound of $n_{str} \geq 7$, as shown in \autoref{tab:minmaxstr} and \autoref{fig:minmaxstr}. For simulation with $128^3$ particles, the median of this peak $n_{str}^{high}(H_i)$ value is $17$. Unsurprisingly, the global maximum of the multistream field ($n_{str} = 2831$) is also a local maximum for one of the haloes. On the other hand, the median of lowest multistream value $n_{str}^{low}(H_i)$ in the haloes is 3 (\autoref{tab:minmaxstr}), which is also the first stage of non-linearity.

\begin{table}
\caption{Local maxima and minima of $n_{str}$ in each of 4492 haloes. The highest $n_{str}$ values in the interior of haloes span over a large range of values. Low values of $n_{str}$ in haloes, which are generally near halo boundaries, have a median of 3.}
\begin{tabular}{|l|r|r|}
\hline
Statistics  &  $n_{str}^{high}(H_i)$ & $n_{str}^{low}(H_i)$ \\ \hline
Minimum     & $7$    & $3$  \\ \hline
Maximum     & $2831$ & $459$      \\ \hline
Median      & $17$   & $3$   \\ \hline
\end{tabular}
\label{tab:minmaxstr}
\end{table}

\begin{figure}
\begin{minipage}[t]{.99\linewidth}
  \centering\includegraphics[width=7.cm]{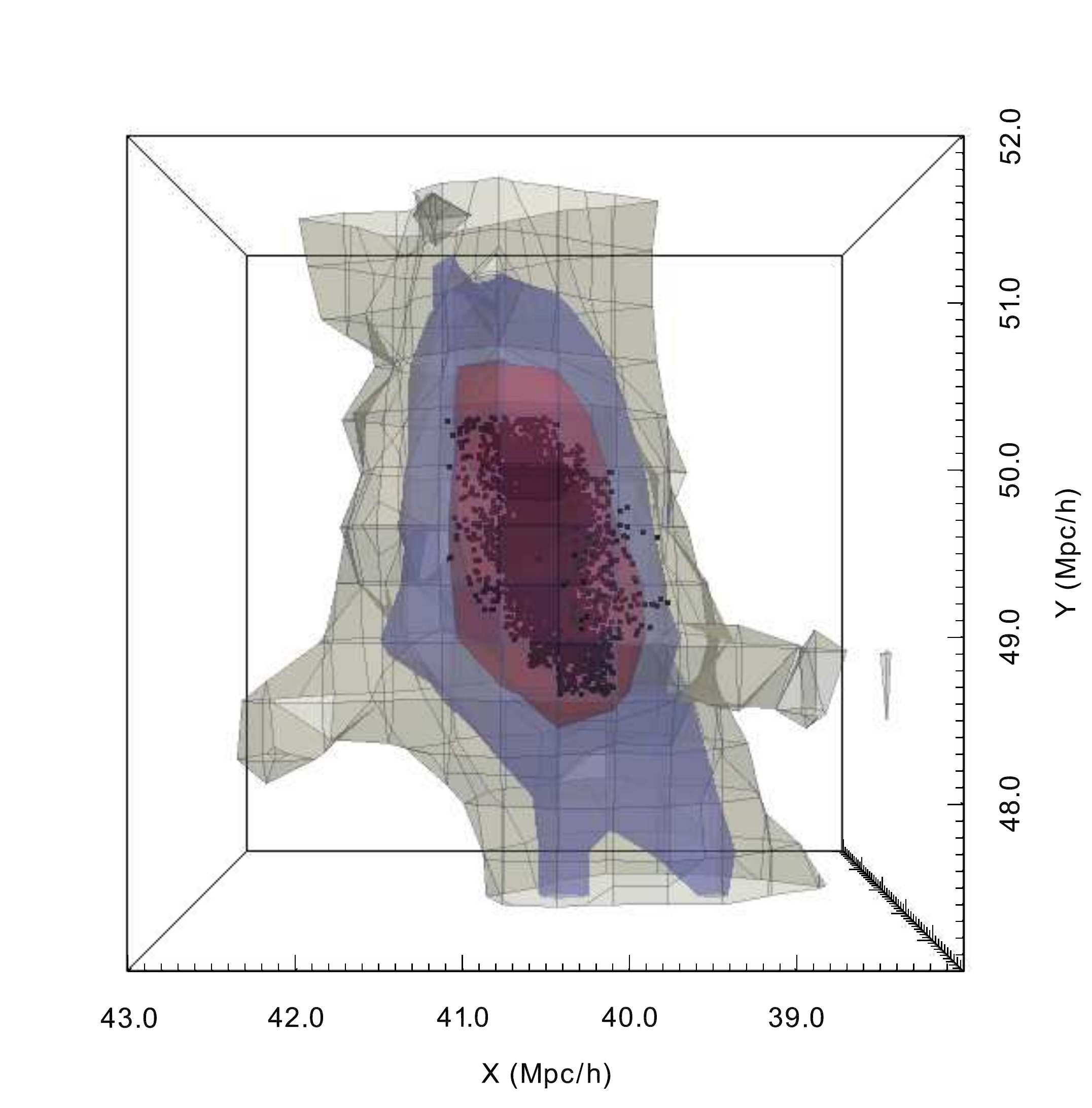} 
\end{minipage}\hfill
\caption{Multistream environment of a $\lambda_3$-halo. The contours represent regions with 3 different multistream values: Outermost $n_{str} \geq 25$ (gray) is tubular, The blue region has $n_{str} \geq 75$. The inner region is highly non-linear with $n_{str} \geq 200$.  The black dots represent the mass particles belonging to a $\lambda_3$-halo, as detected by our algorithm.}
\label{fig:l3Nst}
\end{figure}

An important feature of our halo detection method is that the detected $\lambda_3$-haloes do not have a global threshold on $n_{str}$ or mass density values. The local conditions may be more suited in identifying haloes in multistream fields, since the multistream environments around haloes are very diverse. For instance, regions with $n_{str} \geq 25$ are tubular around one of the the large haloes in \autoref{fig:l3Nst}. Even the region with more than $75$ streams does not enclose a convex multistream region. Whereas, for $n_{str} \geq 200$ the region is convex and the particles detected by our method reside mostly within. We detect closed regions in the multistream field as long as they are not in void, and have at least three foldings in the Lagrangian sub-manifold.

However, the $\lambda_3$-halo boundary is different from any constant multistream contour. That is, from the function $n_{str}(\bmath{x})$, convex regions in the four-dimensional function space $(-n_{str}, x, y, z)$ are projected onto three-dimensional co-ordinate space using eigenvalues. This is different from projecting `iso-multistream' slice onto three-dimensional co-ordinate space. Appendix \ref{appendix:Eigen} illustrates the difference in the two approaches for a one-dimensional function.

The multistream field usually has concentric shells in the regions of successive gravitational collapses (as explained in Section \ref{sec:HaloFormation} and Appendix in \citealt{Ramachandra2017}). In a specific scenario  of \autoref{fig:l3Nst}, regions of lower number-of-streams ($n_{str} = 25$ and lower) is are tubular and have regions of higher $n_{str}$ inside ($n_{str} = 200$ and higher) that is closed. However, this transition from concavity to convexity of the multistream field does not occur at a constant value of $n_{str}$ throughout the field. Instead, it is a local geometrical change that occurs at $\lambda_3 = 0$. For the $\lambda_3$-haloes in our simulation ($N_p = 128^3$), minimum of multistream values $n_{str}^{low}(H_i)$ within each halo has a range of values shown in \autoref{tab:minmaxstr} and \autoref{fig:minmaxstr} -- this varies between $ 3 \leq n_{str}^{low}(H_i) \leq 459$. Hence a global condition on $n_{str}$ does not guarantee that the region is convex.

The particles in a massive $\lambda_3$-halo shown in \autoref{fig:l3Nst} form a spheroidal structure. The number of particles in similar massive haloes are in the order of $10^3 - 10^4$ particles. For instance, the most massive halo in the simulation (with $N_p = 128^3$) has $5593$ particles. We have chosen a minimum threshold of 20 particles, which is an artificial parameter (may be cooked up or ad hoc) used by most halo finder methods. Majority of the $\lambda_3$-haloes have low number of particles; median of number of particles per halo is $49$. Each particle in this simulation is approximately $ 3.65 \times 10^{10} h^{-1} M_{\sun}$. Hence the halo mass range varies in the order of $10^{11} M_{\sun}$ to $10^{14} M_{\sun}$. Combined mass of all the $\lambda_3$-halo candidates is about $31$ per cent of the total mass in the simulation. In contrast, the haloes occupy just $0.3$ per cent of the total volume. Thus the $\lambda_3$-haloes are extremely dense structures. Further analysis of halo mass function of $\lambda_3$-haloes and comparison with AHF- and FOF-haloes is done in the Section \ref{sub:compareHalo}.

\section{Correspondence with other halo finders}
\label{sub:compareHalo}

Comparison of haloes obtained from AHF and FOF method, along with our geometric analysis of the multistream field reveals several interesting features. The number of haloes ($N_H$) with at least 20 particles that were detected by all the algorithms is shown in \autoref{tab:HaloFinderMF}. For both the simulations, FOF detects the highest number of haloes and AHF detects the least. By applying the Hessian algorithm on multistream fields smoothed at the scale of diagnostic grid size, $l_d$, we detected around $4500$ and $28000$ haloes in simulations with $128^3$ and $256^3$ particles respectively. The number of $\lambda_3$-halo is close to the mean of AHF- and FOF- haloes in each simulation -- i.e., $N_H^{\lambda_3}$ is around $2$ per cent of mean of $N_H^{\rm AHF}$ and $N_H^{\rm FOF}$ for the $N_p = 128^3$ simulation and $8$ per cent for the $N_p = 256^3$ simulation. Multistream field both the simulations we calculated at a refinement factor of $l_l/l_d = 2$. 

\begin{table}
  \caption{Number of haloes, $N_H$ detected by the three halo finder algorithms in the two simulations of $L = 100 h^{-1} \text{Mpc}$ with different mass resolutions. Values shown for $\lambda_3$-haloes are done in the multistream fields with refinement factor of 2, and smoothing scale equal to the diagnostic grid size.}
\begin{tabular}{|l|r|r|r|}
\hline
$N_p$  &  $N_H^{\rm AHF}$ & $N_H^{\lambda_3}$ &  $N_H^{\rm FOF}$  \\ \hline
$128^3$   & 3374  & 4492 &  5440  \\ \hline
$256^3$   & 24710  & 27929 & 35765  \\ \hline

\end{tabular}
\label{tab:HaloFinderMF}
\end{table}

\begin{figure}
\begin{minipage}[t]{.99\linewidth}
 \centering\includegraphics[width=8.cm]{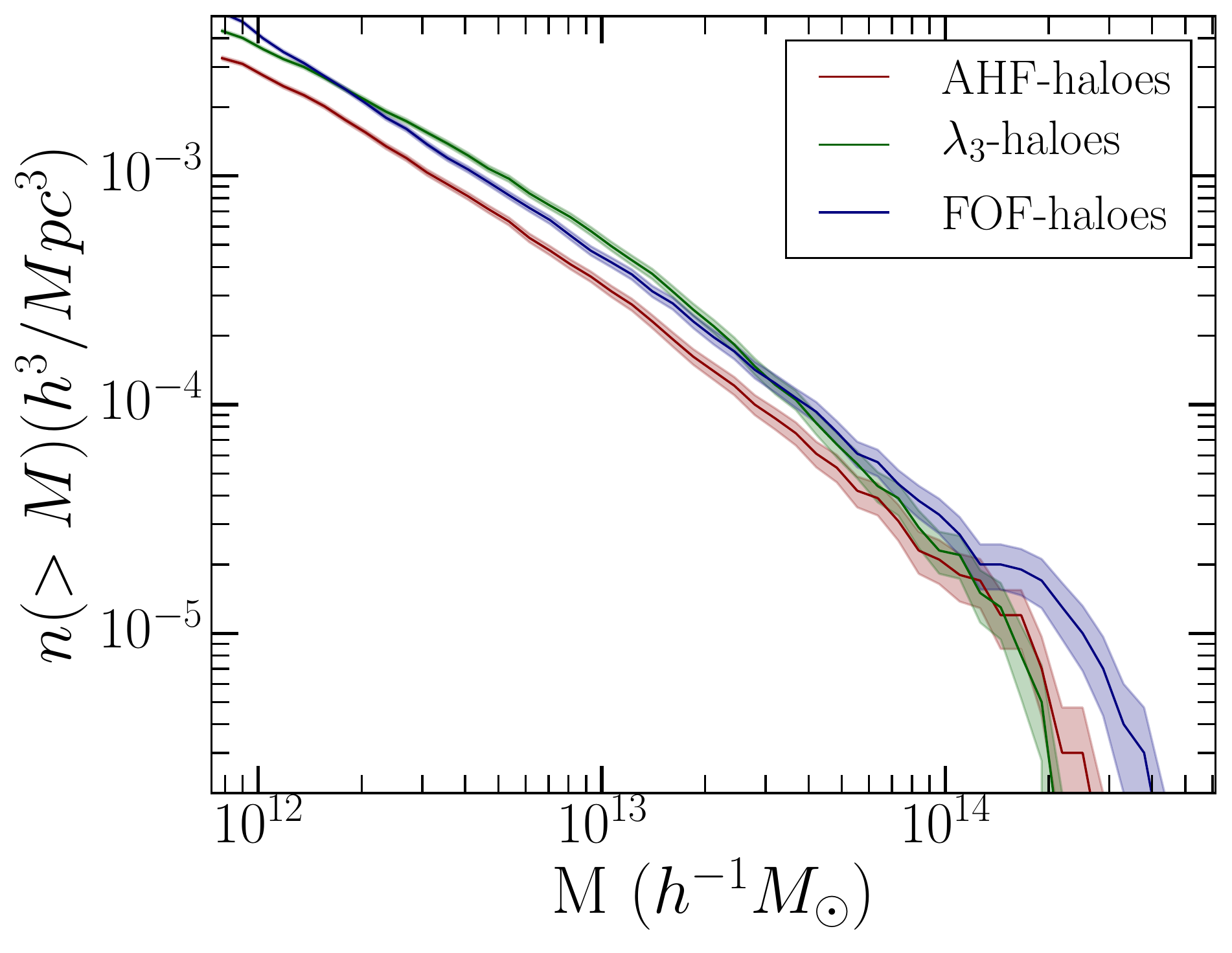} 
\end{minipage}\hfill
\caption{Halo Mass functions from AHF-, FOF- and $\lambda_3$-haloes. The AHF-haloes are fewer than FOF- or $\lambda_3$-haloes. The number of haloes above a mass threshold are binned and taken along vertical axis, normalized to simulation box volume. Error of 1$\sigma$ is shown in shaded region. }
\label{fig:hmf}
\end{figure}

The halo mass functions from all three finders are shown in \autoref{fig:hmf}. For smaller haloes of order of $10^{13} M_{\sun}$, our method predicts a slightly higher number of haloes than FOF and AHF. For the most massive haloes of mass around $10^{14} M_{\sun}$, number of $\lambda_3$-haloes is fewer than the other 2 methods, albeit around the range of error of AHF-haloes.

\begin{figure}
\begin{minipage}[t]{.99\linewidth}
 \centering
 \includegraphics[width=8.cm]{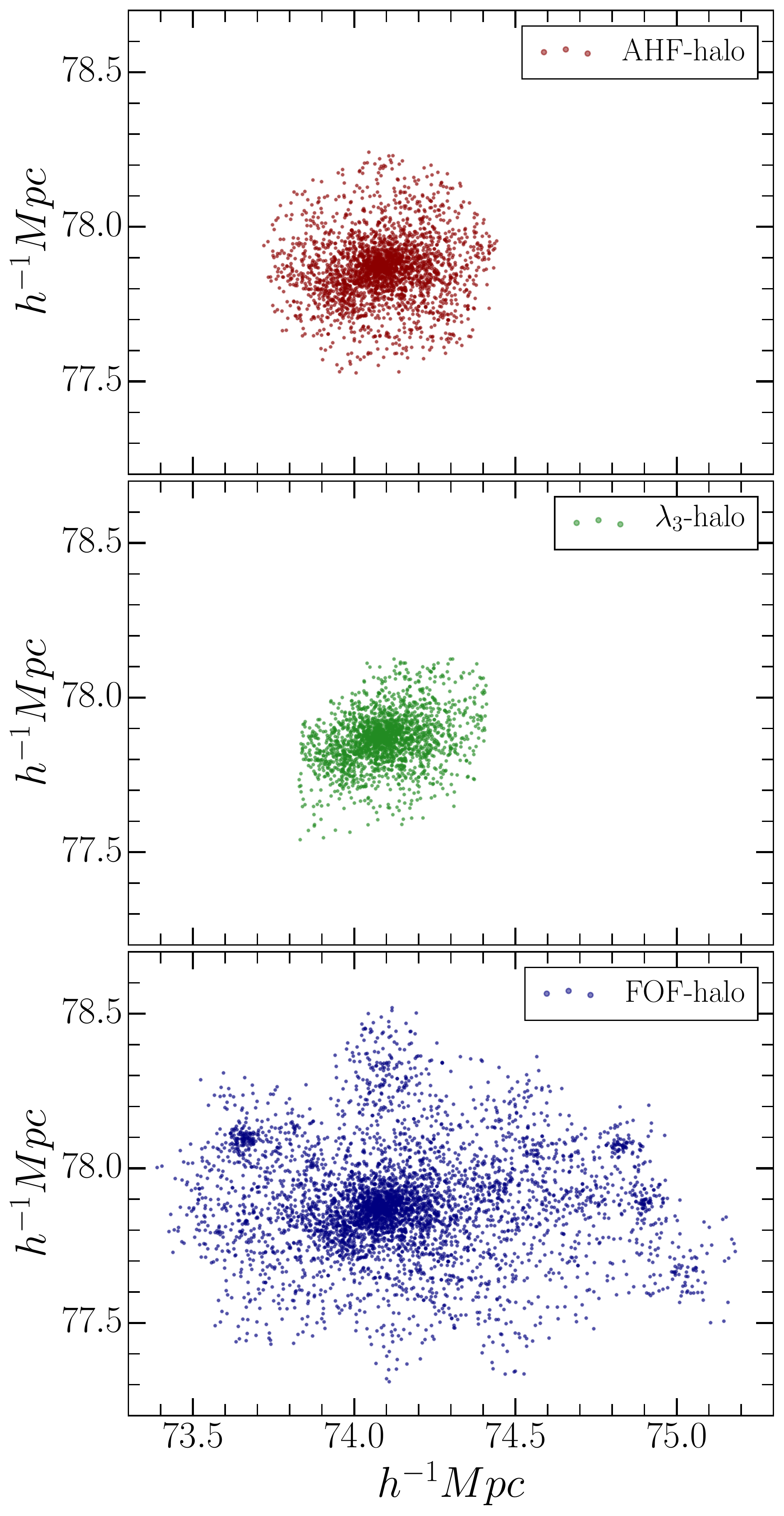}
\end{minipage}\hfill
\caption{A large halo that was detected by AHF (top, red), our geometric analysis in the multistream field (centre, green) and FOF (bottom, blue). Halo boundary differs for each halo finder method. AHF detects particles within a sphere of virial radius. FOF-halo is irregularly shaped. $\lambda_3$-halo particles are in a non-spherical, yet compact structure. }
\label{fig:FinderCompareAll}
\end{figure}

By observing some of the massive haloes, like the one in \autoref{fig:FinderCompareAll}, we find that the $\lambda_3$-halo particles are within AHF- or FOF-halo region. This is generally observed in other massive haloes too: the large $\lambda_3$-haloes have fewer particles than corresponding AHF- or FOF-haloes. For haloes greater than $10^{14} M_{\sun}$, $\lambda_3$-haloes have boundaries slightly within the AHF virial radius. Without unbinding, the FOF-haloes can be very large compared to other methods, as seen in \autoref{fig:FinderCompareAll}. This results in a deviation in the $\lambda_3$-halo mass function (\autoref{fig:hmf}) from the other two methods over halo mass of $10^{14} M_{\sun}$. Further discussion of size of the detected $\lambda_3$-haloes in the context of smoothing of the multistream is done in Section \ref{sub:Smooth}.

\begin{figure}
\centering

\includegraphics[width=7.7cm]{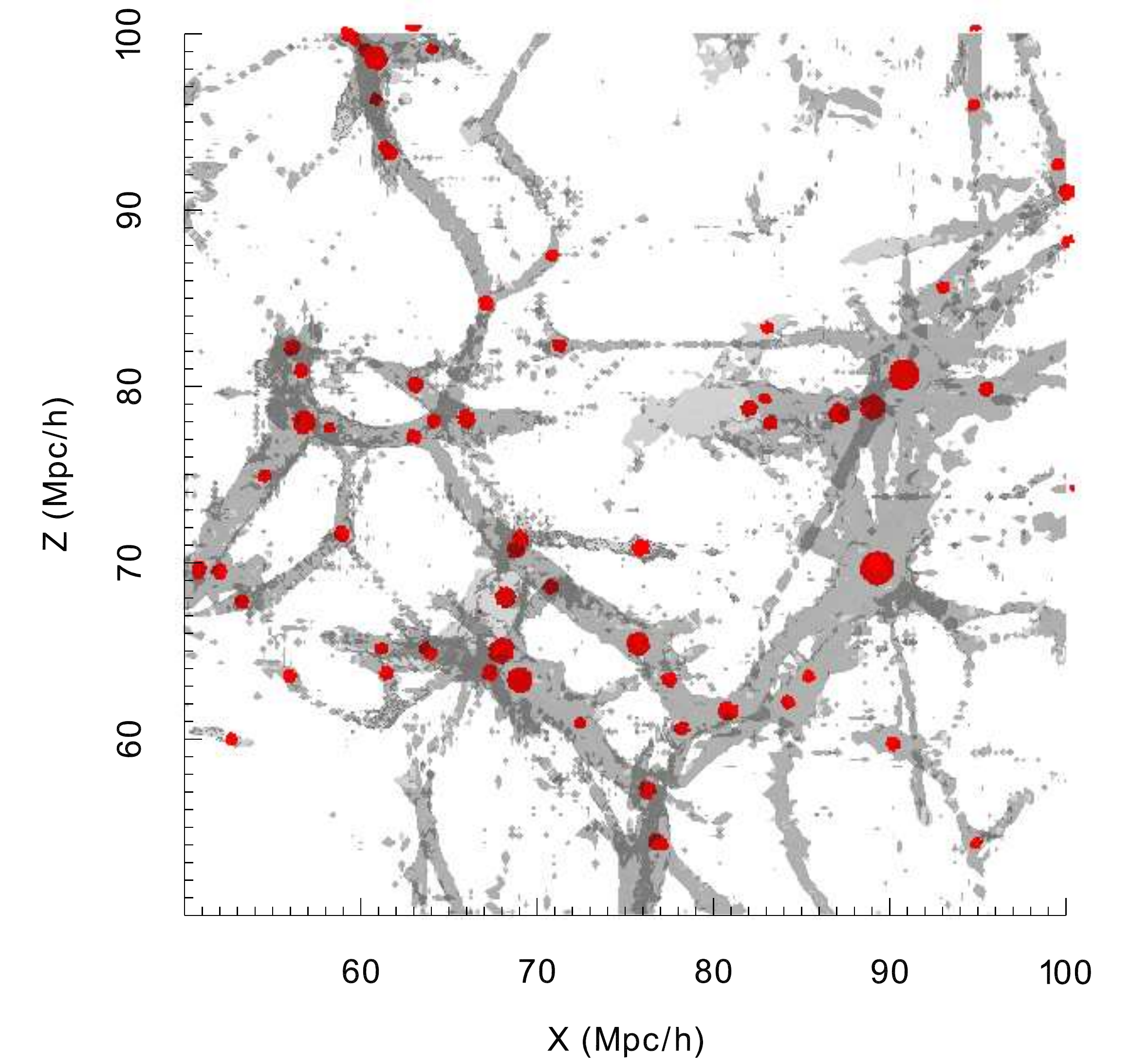}  
\includegraphics[width=7.7cm]{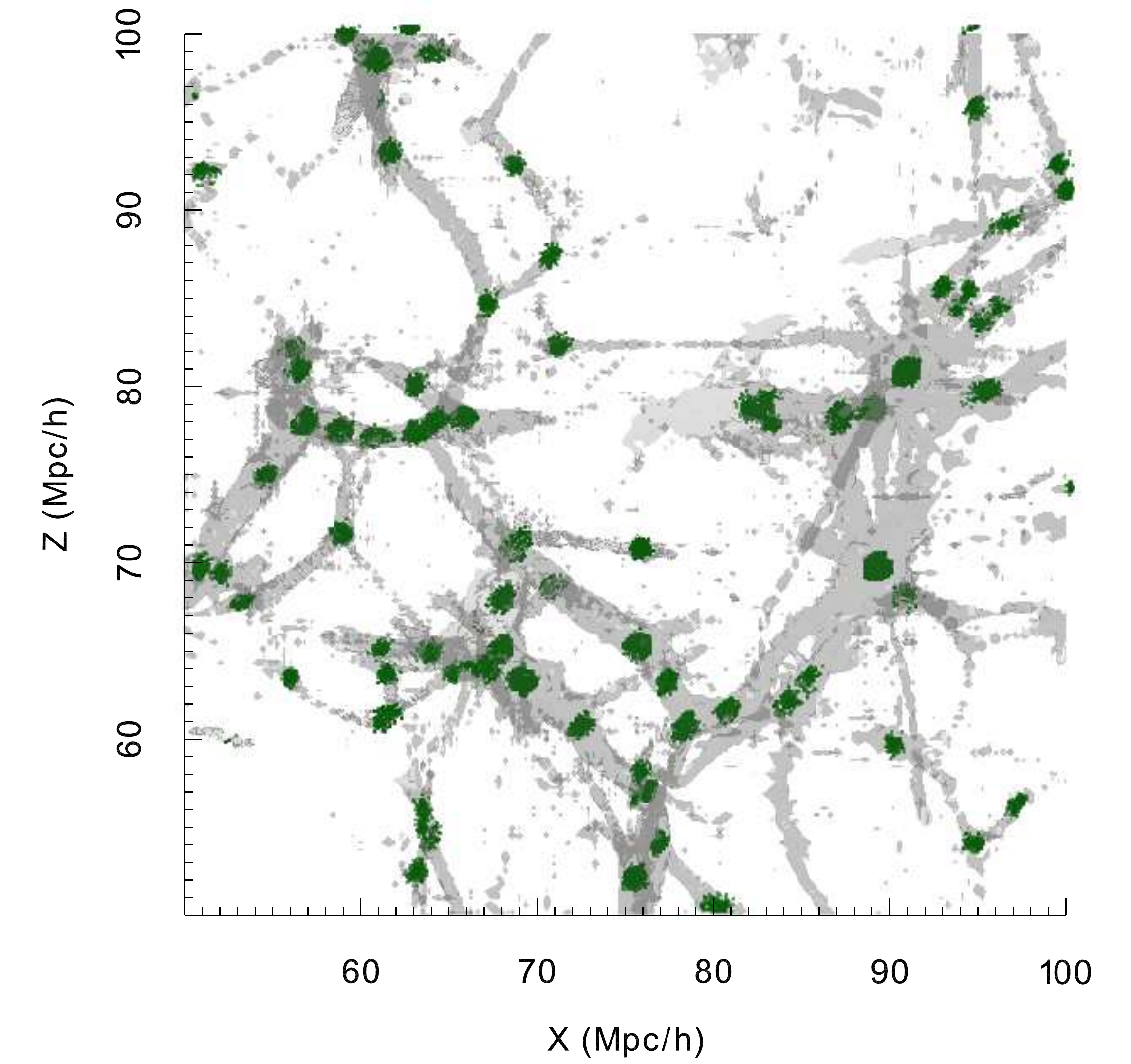}  
\includegraphics[width=7.7cm]{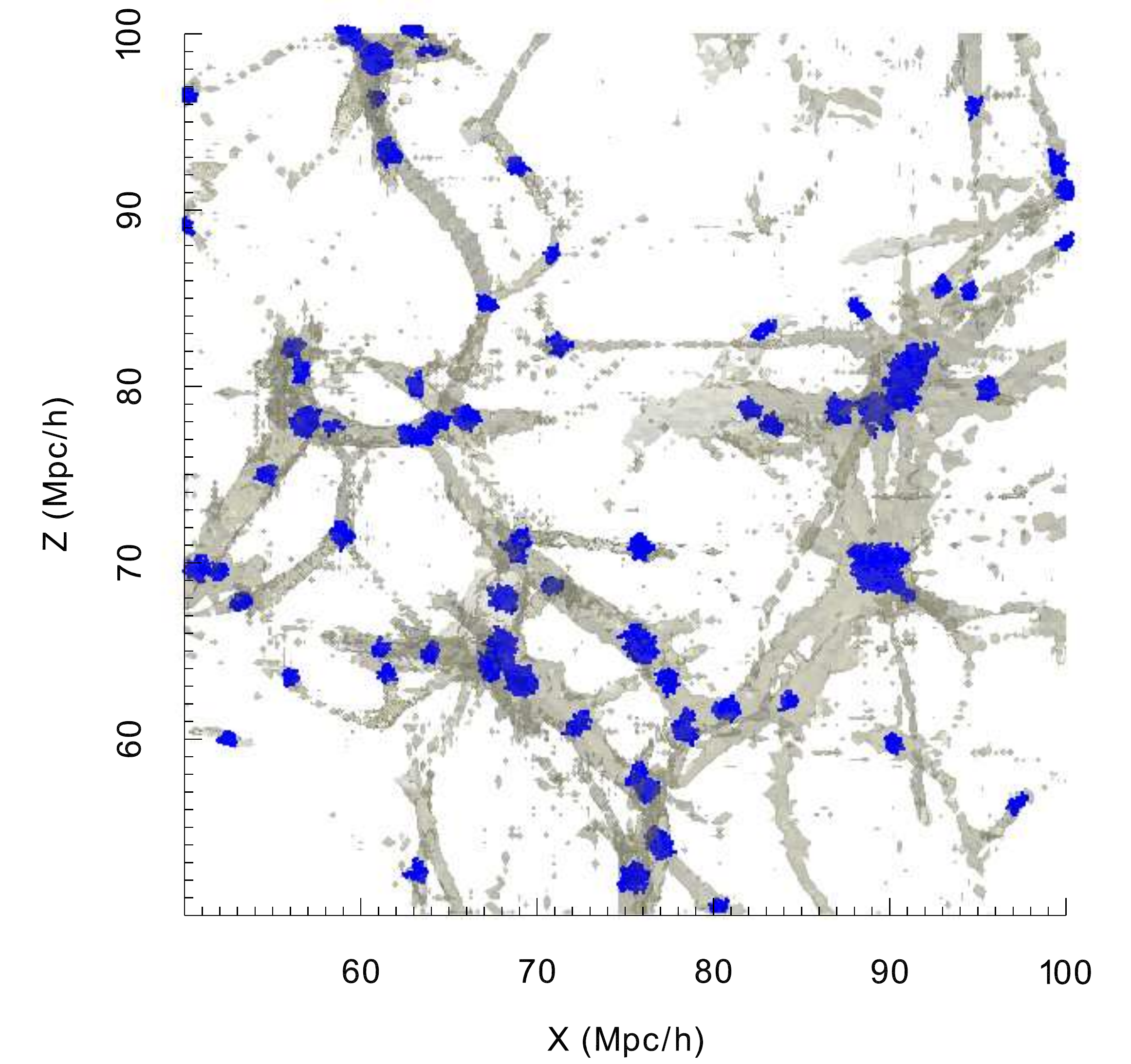} 

\caption{Potential haloes detected by AHF (top), our analysis (centre) and FOF (bottom). Most of the haloes are embedded in a percolating filament with $n_{str} \geq 9$. AHF-haloes are spherical by definition. FOF-haloes without any post processing are elongated along the filament. $\lambda_3$-halo candidates are neither spherical, nor elongated. Boundaries of $\lambda_3$-haloes are well resolved.} 
\label{fig:FinderCompare3}
\end{figure}

The particles identified by the AHF as belonging to haloes form spherical structures due to a series of processes (including virialization) applied to unbind the particles. Inherently, the iso-density contours at virial levels are not spherical or spheroidal. The virialized AHF-haloes on the web are shown in the top panel of \autoref{fig:FinderCompare3}. However, the spherical AHF-haloes are fewer in number compared to the other methods. 

The popular choice of linking length of $b=0.2$, although corresponding to virial density, does not ensure that the haloes have positive curvature. Most algorithms based on the FOF method re-define the halo boundaries by unbinding the particles outside a truncation radius. This truncation radius maybe the distance from the centre of mass of the halo to the farthest particle, rms distance, or an inflection point in the density field (For details on these methods, see \citealt{Knebe2011a} and references therein). Some halo finders define the virial radius, $r_{vir}$ at the distance from halo center where the density is $\Delta_{vir}$ times the background density. In the middle panel of \autoref{fig:FinderCompare3}, the FOF-haloes are shown without any of the above post-processing schemes. Without any unbinding, the FOF-haloes are generally larger in size than $\lambda_3$-haloes in the centre panel of \autoref{fig:FinderCompare3}. For a specific case of a massive halo, \autoref{fig:FinderCompareAll}, FOF identifies more particles as bounded, than AHF or our algorithm.

In contrast to the AHF and FOF algorithms, our halo method directly detects a closed, convex surface (approximately the largest one, since $\lambda_3 >0$) for each of the haloes. There is no unbinding procedure on the particles identified within the halo site. The boundaries of $\lambda_3$-haloes themselves are not spherical or of any regular structure, but they are closed convex surfaces, as seen in the middle panel of \autoref{fig:FinderCompare3}.

Haloes from the three finders in \autoref{fig:FinderCompare3} 
also show other differences in the halo boundaries. While all the AHF-haloes are spherical by definition, the FOF-haloes are irregular. The boundaries of the $\lambda_3$ haloes are not spherical either, but are more compact than FOF-haloes and in addition they are convex surfaces by design. At some junctions of the filaments, FOF identifies a large region as belonging to single halo, whereas AHF and our algorithm detect multiple isolated haloes. Each isolated $\lambda_3$-halo region enclose one maximum of multistream field, thus guaranteeing that multiple haloes are always resolved. On the other hand, a linking length cut-off or a constant threshold on scalar fields may enclose regions with multiple local maxima ( For one-dimensional fields, an illustration of this is shown in Appendix \ref{appendix:Eigen}).

\begin{figure}
\begin{minipage}[t]{.99\linewidth}
 \centering
 \includegraphics[width=8.cm]{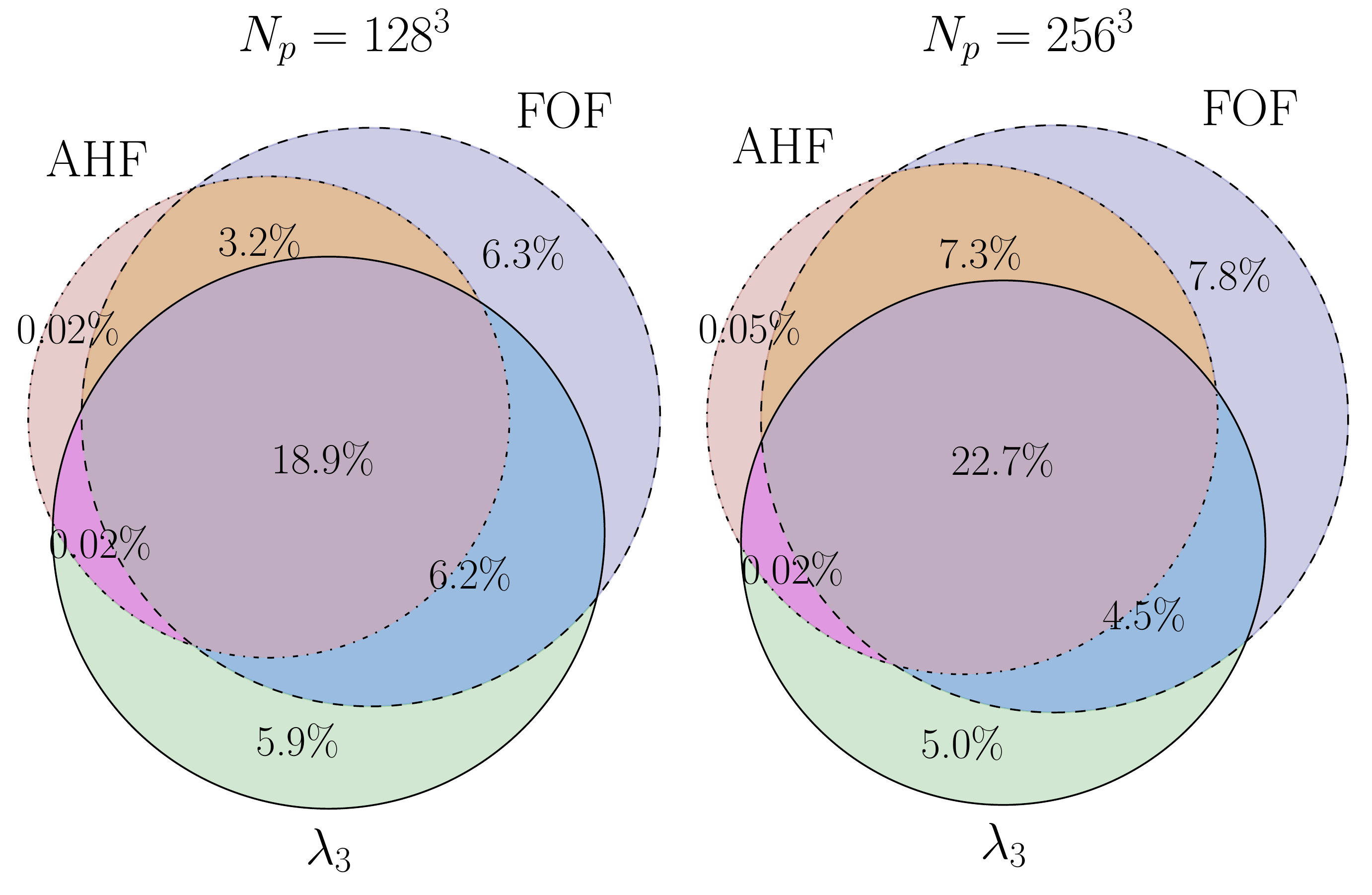}
\end{minipage}\hfill
\caption{Mass fraction of haloes $f_h$ (in per cent) as a detected by the three finders. Each circle represents fractions of mass of all halo particles (AHF, FOF or $\lambda_3$) in the total mass of the simulation box. The particles concurrently detected as belonging to haloes by different frameworks are shown in the intersections. }
\label{fig:HaloFinderMF}
\end{figure}

For a simulation box with $N_p$ particles, each of mass $m_p$, the halo-mass fractions, $\displaystyle f_h = (\sum\limits_{i=1}^{N_H} m_H(i)) / (m_p N_p)$ (where $m_H$ is the mass of each halo and $N_H$ is total number of haloes) are shown in a Venn diagram in \autoref{fig:HaloFinderMF}. For the simulation with $N_p = 128^3$ particles, AHF-, $\lambda_3$-, and FOF- haloes occupy $22$, $31$ and $35$ per cent of the total mass respectively. Nearly $19$ per cent of the total mass are concurrently detected as belonging to haloes by all the three algorithms. FOF (with highest halo mass fraction) detects virtually all the haloes that AHF (with least halo mass fraction). About $3$ per cent of the particles classified as belonging to haloes by both AHF and FOF are not classified as multistream halo particles. Our method also detected nearly $6$ per cent of mass particles as haloes, which neither FOF nor AHF classify as haloes. For simulation with $N_p = 256^3$ particles, the corresponding halo mass fractions $f_h^{\rm AHF} = 30$ per cent, $f_h^{\lambda_3} = 32$ per cent, and $f_h^{\rm FOF} = 42$ per cent respectively. Thus the mass fraction $f_h^{\lambda_3}$ remains fairly consistent over increasing mass resolution, as opposed to AHF and FOF. However, large fractions of these mass particles, nearly $23$ per cent of the $N_p = 256^3$ (increased from $19$ per cent for low mass resolution simulation), are simultaneously detected as belonging to haloes by different methods, as shown in the right panel of \autoref{fig:HaloFinderMF}. For the simulation with $N_p = 256^3$, we also see increase in agreement between any two pairs for halo finders. That is, the mass fraction of haloes simultaneously detected by $\lambda_3$ and FOF jumps from $25$ per cent to $27$ per cent. This correspondence increases from $19$ to $23$ per cent for $\lambda_3$-AHF, and $22$ to $30$ per cent for FOF-AHF pairs. For the same pair (in the $N_p = 256^3$ simulation), $12.3$ per cent of particles are detected by FOF but not AHF, whereas almost all the particles ($>99.9$) for the AHF particles were also detected by FOF. For the pair $\lambda_3$-AHF, $9.5$ per cent of particles are detected by $\lambda_3$ but not AHF, and $7.3$ per cent of particles were detected by AHF but not by $\lambda_3$. Finally, for the $\lambda_3$-FOF pair, $5$ per cent of particles are detected by $\lambda_3$ but not FOF, and $15$ per cent of particles were detected by FOF but not by $\lambda_3$. 

On the other hand, looking at the mass particles that were only detected as haloes by one method, but not by other two, we see that only the multistream haloes improve (i.e., the disagreement reduces from $5.9$ to $5$ per cent) with mass resolution. FOF detects $6.3$ and $7.8$ per cent of haloes in simulations of $128^3$ and $256^3$ particles respectively, that were not classified as haloes by the other two methods. AHF-halo particles, being sub-set of FOF-haloes for the most part, show less than $0.1$ per cent disagreement with other finders.

The discrepancies may have to be addressed on a case-by-case basis. One of the major difference between the haloes detected by isolating local multistream maxima regions and AHF/FOF is shown for a large halo in \autoref{fig:FinderCompareAll}. Without any unbinding procedure, FOF may detect very large irregular sized haloes, often consisting of multiple sub-haloes as shown in the bottom panel. On the other hand, the corresponding AHF-halo (top panel) is smaller spherical subset of FOF-halo. Furthermore, the $\lambda_3$-halo in the middle panel is smaller than both. Our multistream field detection technique selects convex regions with strictly one $n_{str}$ maxima within them. The sub-haloes detected by FOF (or AHF), may be detected as separate $\lambda_3$-haloes. Nevertheless, some of the mass particles between the two neighboring haloes (like ones along saddle regions of multistream fields) will not be included as belonging to the halo. This effect is seen in halo mass functions (\autoref{fig:hmf} for large haloes of mass more than $10^{14} M_{\bigodot}$ -- number density of large $\lambda_3$-haloes is smaller than FOF. Similarly it causes a few discrepancies in mass fractions of potential haloes as well. 

Other cause for differences in mass fraction is also rooted in the definition of haloes. Single-streaming regions are excluded from our halo search completely. Whereas, FOF and AHF employ no such mechanism to check for number of gravitational collapses. \cite{Ramachandra2015} showed that a significant fraction (nearly $35$ per cent) of FOF-haloes are in contact with single-streaming voids. Particles within these regions would not be considered as potential $\lambda_3$-halo particles. This also contributes to the discrepancy in halo mass fraction by different halo finders.

\subsection{Haloes in the percolating filament} 

\begin{figure}
\begin{minipage}[t]{.99\linewidth}
  \centering\includegraphics[width=8.cm]{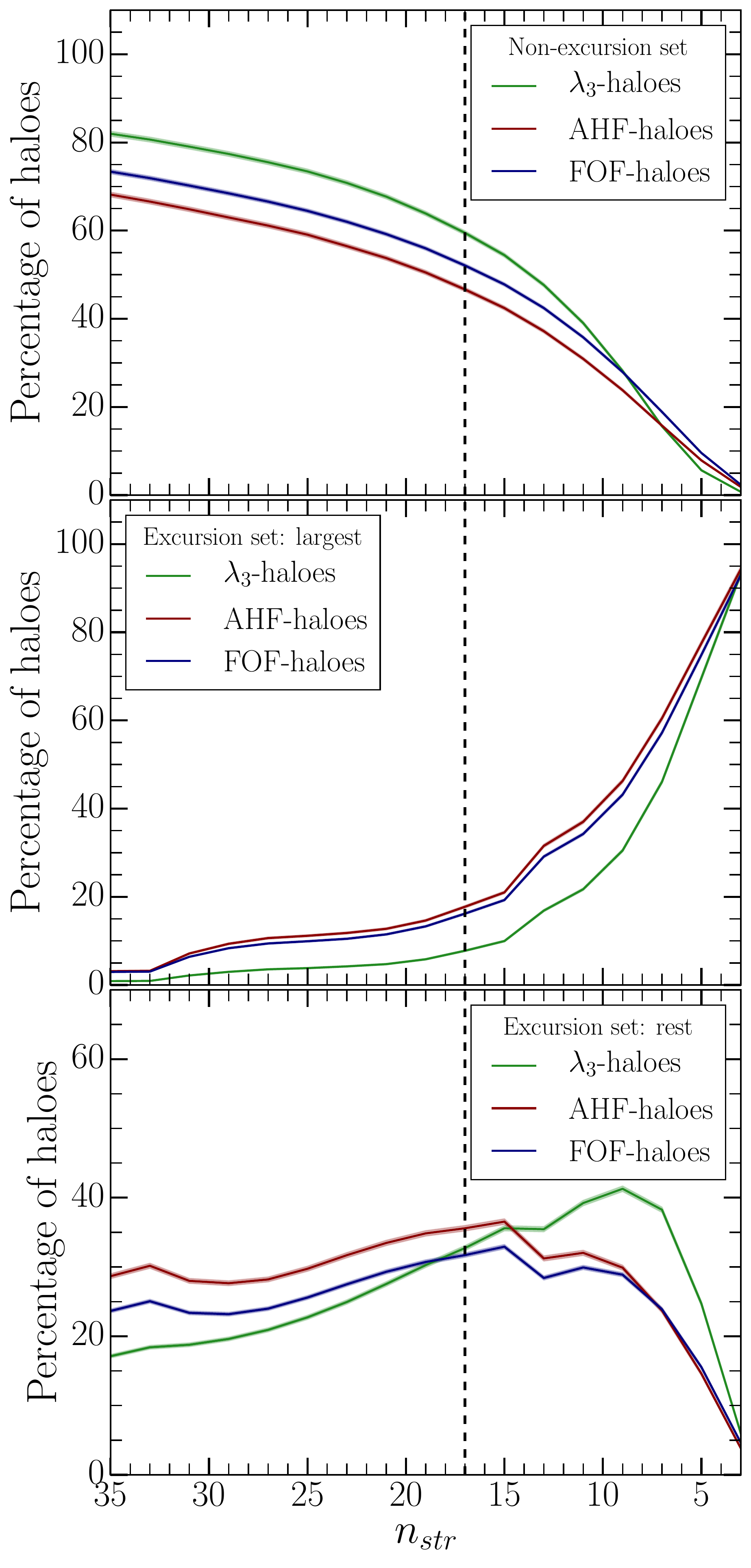} 
\end{minipage}\hfill
\caption{ Percentage of haloes detected (in the simulation with $N_p = 256^3$) that are embedded in the non-excursion set (top panel), largest excursion set segment (center panel) and the rest of excursion set regions (bottom panel). Multistream haloes, AHF-haloes and FOF-haloes are shown in green, red and blue respectively. Vertical dashed line at $n_{str} = 17$ is where percolation transition occurs. }
\label{fig:HaloFilAll}
\end{figure}

The excursion set of multistreams above an $n_{str}$ threshold hosts a varying number of haloes. We compare the multistream halo candidates from our geometric method with the AHF and the FOF method in \autoref{fig:HaloFilAll} 

for the simulation with $N_p = 256^3$. The regions in the co-ordinate space are classified into excursion set and non-excursion set regions based on whether the multistream is over or under the $n_{str}$ threshold. In the excursion set we also distinguish the largest structure from  the rest of the structures because the largest region of the excursion set plays the crucial role  in detecting the  transition to percolation. Percolation takes place at thresholds $n_{str} \le 17$ (\citealt{Ramachandra2017}) to the right from  the vertical dashed dashed line. Based on the coordinates of the halo particles, we check if a halo is in contact with the largest region of the excursion set or with rest of the excursion set.

The fraction of haloes in the non-excursion set are shown at various $n_{str}$ thresholds in the top panel of \autoref{fig:HaloFilAll}. At thresholds greater than 17 streams (i.e. in non-percolating regime), a large fractions of the AHF-, FOF- and $\lambda_3$-haloes are in the non-excursion set, as shown in the top panel of \autoref{fig:HaloFilAll}. The fraction of $\lambda_3$-haloes is slightly higher than FOF or AHF in this regime. At relatively high threshold of, say, $n_{str} = 35$, about 65\% of the AHF-haloes, about 75\% of the FOF-haloes and about 80\% of the $\lambda_3$-haloes are in the non-excursion set.

For multistream values lower than the percolation threshold of $n_{str} \le 17$ (i.e. in percolating regime) the fractions of AHF-, $\lambda_3$-  and FOF-haloes in the largest (i.e. percolating) region quickly grow with decreasing values of the threshold and surpass both the corresponding fractions in the non-percolating  regions of the excursion set and that in the non-excursion set
at $n_{str} \approx 10$. The majority of the haloes belong to the single percolating structure (shown for the simulation with $N_p=128^3$ in \autoref{fig:FinderCompare3}. Similar spatial distribution of SUBFIND haloes \citealt{Springel2001a} in the multistream regions is shown by \citealt{Aragon-Calvo2016}) and at $n_{str} = 3$, all the haloes are attached to the web.  

At $n_{str} = 3$, the filling fraction $f_1 / f_{ES}$ is almost close to unity \cite{Ramachandra2017}. Most halo candidates from all three algorithms are at least in contact with the percolating cosmic web. Due to the threshold on $n_{str}$ in our halo detection method, the $\lambda_3$-haloes are not only in contact, but completely within this structure.

\section{Discussion}
\label{sec:discussion}

The multistream field can have only integral  values, moreover these values must be odd numbers practically at every point, because the number of streams can be even only on a set of measure zero. It means that in numerical simulations  even values may occur on extremely rare occasions. 
We have analysed functional variation of the scalar field $-n_{str}(\bmath{x})$ using Hessian eigenvalues. The Hessian analysis is generally done for inherently continuous fields since it requires  evaluation of the second derivatives. 
Geometries of web structures unveiled by Hessian signatures of smoothed density fields (such as \citealt{Sousbie2008a}, \citealt{Aragon-Calvo2007}, \citealt{Aragon-Calvo2010a}, \citealt{Cautun2014a}, \citealt{Bond2010a} and many more), tidal shear or velocity shear tensor (\citealt{Hahn2007}, \citealt{Forero-Romero2009a}, \citealt{Hoffman2012a}, \citealt{Hoffman2012a}, \citealt{Libeskind2013}, \citealt{Cautun2014a} etc.) and observational data from surveys (\citealt{Sousbie2008a}, \citealt{Bond2010a}, \citealt{Bond2010b}, \citealt{Pahwa2016} etc.).

Although the multistream field has discrete values by definition, it may be smoothed for numerical analysis at some scale, typically the scale of grid length of the field. The resulting Hessian eigenvalues approximately characterize the geometry in a four-dimensional hyper-space of ($-n_{str}, x, y, z$). Our only assumption about  the shape of the boundary of a virialized halo is that it is a convex surface.  Therefore the boundary of a halo can be defined as a region with $\lambda_1 \geq \lambda_2 \geq \lambda_3 > 0$ since it is a closed convex contour in  the ($-n_{str}, x, y, z$) hyper-space, and thus it's projection onto the three-dimensional  $(x,y,z)$ space is also closed and convex. 

Differentiating a smoothed  $n_{str}(\bmath{x})$-field may still pose a problem in the regions where $n_{str}(\bmath{x})$ is close to a constant and therefore the eigenvalues represent noise about the zero level. Fortunately in the most of such regions the unsmoothed $n_{str}(\bmath{x}) =1$ therefore they can be easily eliminated.

Our algorithm for detecting potential dark matter haloes is unique due to two important factors: the geometrical attribute and the choice of field. Local geometrical analysis on the multistream field conveniently delineates the non-void structures without any free parameters. The dark halo candidates have compact surfaces that enclose local maxima of the multistream field. We do not employ non-local thresholds that several halo finders use (see \citealt{Knebe2011a}, \citealt{Knebe2013} and \citealt{Onions2012} for comparisons of various halo finders). Global thresholds (like a constant $n_{str}$ cut-off) might be unsuitable for detecting halo candidates since the halo multistream environments are generally not categorical. Secondly, the $n_{str}$ field enables us to mask out the regions belonging to mono-stream regions without a heuristic criteria. Our method guarantees that none of the $\lambda_3$-halo particles belong to void region.

We note that present halo finders employ a variety of physical and numerical processes to identify dark matter halo candidates. Furthermore, there is no consensus in the definition of haloes itself (see discussion in \citealt{Knebe2011a}). This is also the cause for the few differences between FOF-, AHF- and multistream haloes: FOF and AHF haloes only use Eulerian co-ordinates $\mathbf{x}(z)$ -- either raw positions or in the form of mass density. On the other hand, we utilize a mapping on the Lagrangian sub-manifold $\mathbf{x}(\mathbf{q}, z)$ to define the multistream field $n_{str}(\mathbf{x})$. The boundaries of haloes in FOF and AHF are defined by the free-parameter thresholds of linking length and density, and the halo-center is usually defined as the center of mass of the particles. Conceptually, the center of a $\lambda_3$-halo is the location of the local multistream maximum, and the boundary of the halo is the convex region surrounding it. At least for large haloes like the one in \autoref{fig:FinderCompareAll}, this convex region is well within the FOF boundary.  

The applicability of non-local thresholds in detecting haloes deserves deeper investigation. Lower bounds on over-density or linking-length thresholds traditionally define halo regions in several halo finders. Values such as $\Delta_{vir} \approx 180$ or $b \approx 0.2$ correspond to virial theorem applied to isolated spherical collapse models. Recently \cite{More2011} demonstrated that depending on halo environment, cosmology and redshift the over-densities corresponding to $b = 0.2$ have different values. The virial theorem itself is a good measure of equilibrium of a system. However, the global thresholds empirically derived from it may not be pertinent to diverse environment of dark matter haloes.

The algorithm prescribed in Section \ref{sec:haloDetection} lists out a set of physically motivated steps that filter out the noisy $\lambda_3 > 0$ regions that cannot be identified as haloes. The analysis in the simulation of $100 h^{-1}$ Mpc side length and $128^3$ particles, with the multistream calculated on $256^3$ diagnosis box, approximately 40000 labelled segments satisfying $\lambda_3 > 0$ criterion in the non-voids were filtered out by a lower limit on multistreaming regions. One of the possible improvements in our algorithm would be to use information of number of flip-flops of each particle (For instance, using methods prescribed by \citealt{Shandarin2014,Shandarin2016}). Such methods involving the Lagrangian sub-manifold may reveal rich sub-structure in the haloes.

The requirement that each halo should have closed convex surfaces with a multistream maximum inside may identify sub-haloes in large haloes but is too demanding because a halo with sub-haloes must have saddle points in the $n_{str}$ field. This may explain the shortage of massive haloes shown in \autoref{fig:hmf}. Applying a filter for smoothing the $n_{str}$ field increases the number of massive haloes but reduces the number of low mass haloes. Although our present method does not currently perform an analysis simultaneously on multiple smoothing scales, such approaches done in density, lognormal density, tidal, velocity divergence or velocity shear fields (see MMF by \citealt{Aragon-Calvo2007} and NEXUS+ by \citealt{Cautun2013}) have shown interesting multi-scale features of the cosmic web. Applying a more sophisticated procedure for linking $\lambda_3$-sub-haloes in a more massive halo will be done in the follow-up paper.

Dark matter haloes, being localised structures, are uniquely convenient for our local Hessian analysis. Conditions of $\lambda_1 > 0 > \lambda_2 \geq \lambda_3 $ and $\lambda_1 \geq \lambda_2 > 0 > \lambda_3 $ also give information about curvature. Hessian eigenvalue analysis at high resolution of multistream fields may be very interesting in understanding the tubular edges of filaments and surfaces of walls at smaller scales. However, in this study, Hessian analysis is only applied to haloes. Walls and filaments span large volumes in the dark matter simulations, and we employ topological tools to investigate them.

\section{Summary}
\label{sec:summary}

We studied certain geometrical of the multistream field in the context of halo formation. Findings from our analysis are summarized as follows: 

\begin{enumerate}

\item Several aspects of halo formation in the Lagrangian sub-manifold are considerably different than that of reference models of spherical top-hat collapse and ellipsoidal collapse. Successive formations of caustics (and consequently multiple velocity streams) play a crucial role in the process of clustering.

\item We present a novel halo detection algorithm for identifying dark matter halo candidates in the multistream field. Conditions on the local geometric indicators of the field are used to ensure that each closed halo boundary hosts a local multistream maxima. The 
positive signs of all principal curvatures  (please note that we use curvatures of $-n_{str}(\bmath{x})$ field) inside the boundary also guarantee that the boundary is convex. 
Bounds on $n_{str}$  guarantee that all the halo particles are in the non-void structure. We also ensure that the halo regions have foldings in the Lagrangian sub-manifold in more than one direction.  

\item The multistream field within the halo boundaries may be very diverse. We do not detect halo candidates from a global lower bound on $n_{str}$. Instead, we look for closed convex regions in the multistream field. For the simulation with $128^3$ particles, minima of $n_{str}$ in each halo vary from 3 to nearly 450. Maxima of $n_{str}$ in the halo vary from 7 to about 2800. 

\item Our multistream halo candidates had a reasonably good correspondence with haloes from AHF and FOF catalogues. One notable difference was found with massive haloes. Our algorithm predicted fewer particles than the FOF method. This is likely to be caused
by the requirement that the multistream field in the regions of the $\lambda_3$-halo candidates is convex  which may be a reasonable approximation  for simple haloes (i.e. having no sub-haloes) but massive haloes are more likely to have sub-haloes and therefore the $n_{str}$ field in the corresponding regions must have saddle points ant therefore cannot be entirely convex. Our study of the smoothing effects  has shown that the number of massive sub-haloes tend to increase with growing smoothing scale which seems to agree with the above explanation. We will address this problem in the following study.

\item Halo candidates were mostly embedded on the excursion set of the multistream field after percolation transition ($n_{str} = 17$ in the simulation with $256^3$ particles). At lower thresholds (around $n_{str} = 5 \text{ to } 11$), the largest percolating structure in the excursion set hosts most of the haloes.     

\end{enumerate}

In conclusion, the Lagrangian sub-manifold contains dynamical information of structure formation. We analysed the multistream field that contains the information of foldings in the sub-manifold. In addition, we demonstrated the use of geometrical features of the multistream field in identifying potential dark matter halo candidates in cosmological N-body simulations.

\section*{Acknowledgements}

This work has been funded in part by the University of Kansas FY 2017 Competition General Research Fund, GRF Award 2301155. This research used resources of the Advanced Computing Facility, which is part of the Center for Research Computing at the University of Kansas. We thank Mark Neyrinck and Mikhail Medvedev for discussions and suggestions. We also thank the anonymous referee for insightful comments on improving this manuscript and the editor for help with the submission. 

\bibliographystyle{mnras}
\bibliography{library}

\appendix

\section{Hessian signatures of the multistream field}
\label{appendix:Eigen}

\begin{figure}
\begin{minipage}[t]{.99\linewidth}
  \centering\includegraphics[width=8.cm]{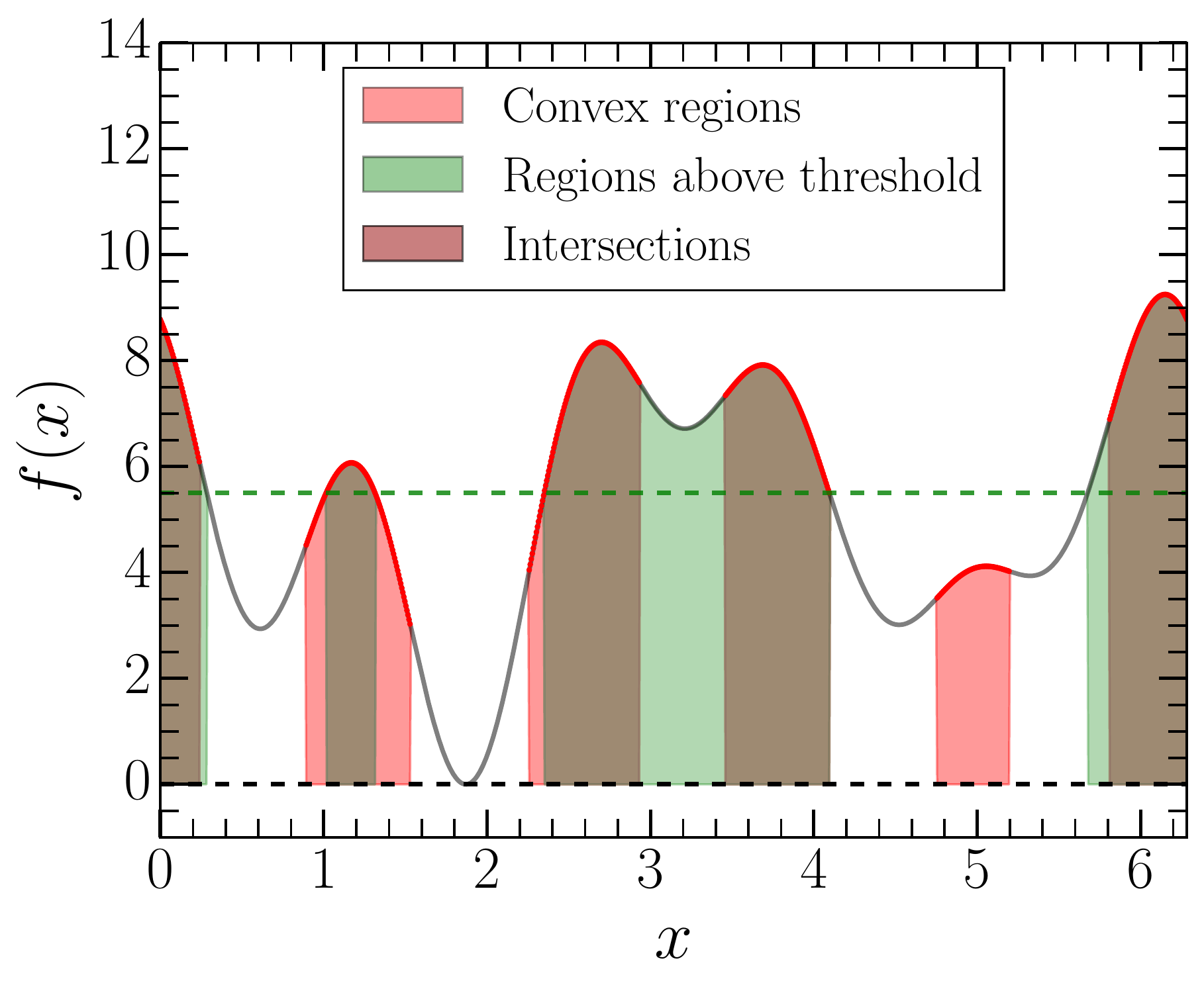} 
\end{minipage}\hfill
\caption{Projections of regions of $f(x)$ from (1+1)-dimensional function space onto one-dimensional coordinate space. Convex regions and regions above a threshold of an arbitrary function $f(x)$ are shown. Both the regions intersect around a few maxima, but not universally.}
\label{fig:check1d}
\end{figure}

Second-order local variations of a scalar field $f$ is described by a Hessian matrix, whose element in a three-dimensional domain is given by \autoref{eq:Hess}. The geometry of the scalar field is classified by the Eigenvalues of the Hessian. The convex regions have at-most one maxima within the (3+1)-dimensional functional space. Projection of this closed region onto three-dimensional coordinate space also gives a closed surface in coordinate space. An illustration of the projection is shown in \autoref{fig:check1d} for a simpler function $f(x)$ in one-dimensional domain. The eigenvalue criteria for regions are simplified: for instance, $\frac{\partial^2 f}{\partial x^2} < 0$ for convex region. Projection of these regions onto coordinate space is shown in the shaded regions. This is different from regions within a contour, which is the projection of the curve along which the function has a constant value. Boundaries of these two regions may, but not necessarily, intersect. 

In the case of cosmic fields, thresholds like $\Delta_{vir}$ are equivalent to the green dotted line in \autoref{fig:check1d}. The over-dense regions (green shaded regions) are not constrained to be convex. Similarly structures selected based on $n_{str}$ thresholds do not universally result in convex structures either. Local geometry can be probed from the eigenvalue criteria instead, as shown by the red line on the curve and corresponding shaded area. The projected structures, albeit convex, may have very small values of $f(x)$ (like the red shaded area around $x=5$). In the framework of identifying potential haloes in multistream field, multistream thresholds are devised in so that some of these small peaks detected by the Hessian are not considered as potential halo sites.

\label{lastpage}

\end{document}